\documentclass[%
aip,
amsmath,amssymb,
reprint,%
]{revtex4-1}
\usepackage{xcolor}
\usepackage{graphicx}
\usepackage{dcolumn}
\usepackage{bm}

\usepackage[utf8]{inputenc}
\usepackage[T1]{fontenc}
\usepackage{mathptmx}
\usepackage{subcaption}
\usepackage{ifthen}
\begin{document}
	
	\preprint{AIP/123-QED}
	
	\title{The Fermi-L\"owdin self-interaction correction for ionization energies of organic molecules}
	
	\author{Santosh Adhikari}
	\email{tuf60388@temple.edu.}
	\author{Biswajit Santra}
	\author{Shiqi Ruan}
	\author{Puskar Bhattarai}
	\author{Niraj K. Nepal}
	\affiliation{Department of Physics, Temple University, Philadelphia, PA-19122, USA}
	\author{Koblar A. Jackson}%
	\affiliation{Department of Physics and Science of Advanced Materials Program, Central Michigan University, Mount Pleasant, MI USA}
	
	\author{Adrienn Ruzsinszky}
	\affiliation{Department of Physics, Temple University, Philadelphia, PA-19122, USA}

	\date{\today}

\begin{abstract}
(Semi)-local density functional approximations (DFAs) suffer from self-interaction error (SIE). When the first ionization energy (IE) is computed as the negative of the highest-occupied orbital (HO) eigenvalue, DFAs notoriously underestimate them compared to quasi-particle calculations. The inaccuracy for the HO is attributed to SIE inherent in DFAs. We assessed the IE based on Perdew-Zunger self-interaction corrections on 14 small to moderate-sized organic molecules relevant in organic electronics and polymer donor materials. Though self-interaction corrected DFAs were found to significantly improve the IE relative to the uncorrected DFAs, they overestimate. However, when the self-interaction correction is interiorly scaled using a function of the iso-orbital indicator z$_{\sigma}$, only the regions where SIE is significant get a correction. We discuss these approaches and show how these methods significantly improve the description of the HO eigenvalue for the organic molecules.
\end{abstract}

\maketitle 
\section{Introduction}
Density functional theory (DFT) \cite{hohenberg1964inhomogeneous,kohn1965self} is widely applied for electronic structure calculations in diverse fields including physics and chemistry. \cite{jones2015density} Though the theory is exact, its practicality often relies on the approximation of the small yet significant quantity called the exchange-correlation (XC) energy (\textit{E$_{XC}$}) via a density functional approximation (DFA).
Depending on its ingredients, there are various forms of such approximations classified as different rungs of the Jacob's ladder of DFT.\cite{jacobsladder2001} The local density approximation (LDA),\cite{kohn1965self,perdew1992accurate} generalized gradient approximation (GGA)\cite{perdew1992pair,pbe1996,perdew2008restoring} and the meta-generalized gradient approximation (MGGA)\cite{staroverov2003comparative,constantin2013meta,scan2015} respectively form the first three rungs of that ladder and are collectively called the semilocal approximations. Although DFAs in general become more accurate (and also more computationally demanding), when moving up the ladder from LDA to MGGA, they all suffer from self-interaction error (SIE) whose origin lies in the inability of the self-exchange-correlation energy (\textit{E$_{XC}$}) to counter-balance the self-Hartree energy ($U[n]$) (shown in Eq. (1)) for a single electron density (as shown in Eq. (2)). \begin{equation}
U[n]=\frac{1}{2} \int \int \frac{n(\textbf{r})n(\textbf{r}^{'})}{|\textbf{r}-\textbf{r}^{'}|}d\textbf{r}d\textbf{r}^{'}
\end{equation}
\begin{equation}
U[n_{i\sigma}] + E_{XC}[n_{i\sigma},0] \ne 0 
\end{equation}
where $\small{n_{i\sigma}=\mid \psi_{i\sigma}\mid^{2}}$, and $\psi_{i\sigma}$ is the "i"th occupied Kohn-Sham (KS) orbital of spin $\sigma$.
Mixing a portion of the exact exchange with the semilocal exchange leads to the fourth rung of the Jacob's ladder popularly known as hybrid functionals.\cite{becke1993new,adamo1999toward,heyd2003hybrid} While such an attempt helps, it does not completely eliminate the SIE.\\
$~~~~$Perdew and Zunger \cite{perdew1981self} found a way to remove the SIE (known as PZ-SIC) through the subtraction of the spurious self-Hartree energy and exchange-correlation terms in an orbital-by-orbital manner from the total energy, thus making the functional exact for all one-electron systems.The exchange-correlation energy in PZ-SIC is
\begin{equation}	
\small{E^{SIC}_{XC}[n_{\uparrow},n_{\downarrow}]=E_{XC}[n_{\uparrow},n_{\downarrow}]-\sum_{\sigma}\sum_{i}^{N_{\sigma}} \left\{ U[n_{i\sigma}] + E_{XC}[n_{i\sigma},0]\right\},}
\end{equation}
where $N_{\sigma}$ is the number of occupied orbitals of spin ${\sigma}$.\\
The resulting functional significantly improves non-equilibrium properties like barrier heights for chemical reactions, dissociation curves of radical molecules, etc., but deteriorites atomization energies when used with semi-local functionals and predicts too short bond lengths in molecules.\cite{goedecker1997critical,csonka1998inclusion} This behavior is referred to as the paradox of SIC.\cite{perdew2015paradox} Since PZ-SIC, by construction, is explicitly dependent on the orbitals, it is not invariant under unitary transformation of the occupied orbitals. Hence, it is computationally demanding to search over all the possible orbitals that yield the correct density for the set that also minimizes the total energy. Furthermore the use of delocalized KS orbitals in Eq. (3) can lead to a breakdown of size-extensivity,\cite{perdew1990size} the principle that the total energy of a system should be equal to the sum of the total energies of its separated systems.\\
Recently Pederson \textit{et al.}\cite{pederson2014communication} introduced the Fermi-L\"owdin-orbitals (FLO) and proposed a localization scheme  as a remedy for some of the aforementioned problems. This method (FLOSIC), whose fully-self-consistent version was achieved by Yang \textit{et al.},\cite{yang2017full} allows one to obtain the orbitals directly from the Fermi orbitals (shown in Eq. (4)) constructed for the unitarily invariant density matrix.
\begin{equation}
\phi_{i\sigma}^{FO}=\frac{n_{\sigma}(\textbf{a}_{i\sigma},\textbf{r})}{\sqrt{n_{\sigma}(\textbf{a}_{i\sigma})}}
\end{equation}
where $n_{\sigma}(\textbf{a}_{i\sigma},\textbf{r})$=$\sum_{j}^{N_{\sigma}}\psi_{j\sigma}^{*}(\textbf{a}_{i\sigma})\psi_{j\sigma}(\textbf{r})$ is the single-particle density matrix of the KS system and $\textbf{a}_{i\sigma}$ is the Fermi orbital descriptor (FOD).\cite{pederson2014communication,yang2017full} More details on FODs can be found in the work of Schwalbe \textit{et al.}\cite{21} Apart from being size consistent and localized, the FLOs are also weakly noded.\cite{shahi2019stretched} These features make real FLOs superior over the delocalized and often highly noded KS orbitals,\cite{shahi2019stretched} with the possibility of further improvement through the use of nodeless complex orbitals.\cite{klupfel2011importance}\\
$~~$The FLOSIC method has been applied to various systems,\cite{pederson2015fermi,sharkas2018shrinking,joshi2018fermi,schwalbe2018fermi,shahi2019stretched,withanage2019self,johnson2019effect,yamamoto2019fermi,trepte2019analytic,sharkas2020self} providing self-interaction corrections to LDA, PBE\cite{pbe1996} and SCAN\cite{scan2015} which we will call LDA-SIC, PBE-SIC and SCAN-SIC, respectively. It has mostly been a success story so far even yielding results comparable to hybrid functionals.\cite{sharkas2018shrinking,joshi2018fermi,withanage2019self} \\   
$~~~~$The ionization energy is one of the few significant properties relevant to the selection of materials in photovoltaics and organic electronics.\cite{korzdorfer2009trust}
High level calculations like CCSD(T), often touted as a gold standard of quantum chemistry, would provide fairly accurate ionization energies,\cite{deleuze2003benchmark,bruneval2013benchmarking,krause2015coupled} but are computationally demanding. As an alternative, Hedin's GW approximation,\cite{hedin1965new} where G and W respectively represent the single particle Green's function and screened Coulomb interaction, is recently getting more attention for molecules.\cite{blase2011first,knight2016accurate,marom2012benchmark,van2015gw,rangel2016evaluating,rostgaard2010fully,bruneval2013benchmarking,faber2011first} There is a hierarchy in GW approximations from a non-self-consistent (G$_{0}$W$_{0}$) version with different mean-field starting points to full-self-consistency, but there is no unambiguous connection to accuracy within this hierarchy.\cite{marom2012benchmark,blase2011first,knight2016accurate,bruneval2013benchmarking} Nevertheless, recent works demonstrate the potential of G$_{0}$W$_{0}$ methods, albeit sensitive to starting points,  performing close to the level of CCSD(T).\cite{rangel2016evaluating,bruneval2013benchmarking}\\
\begin{figure}[h!]
	\captionsetup[subfigure]{labelformat=empty}
	\begin{subfigure}[b]{0.15\textwidth}
		\includegraphics[scale=0.05]{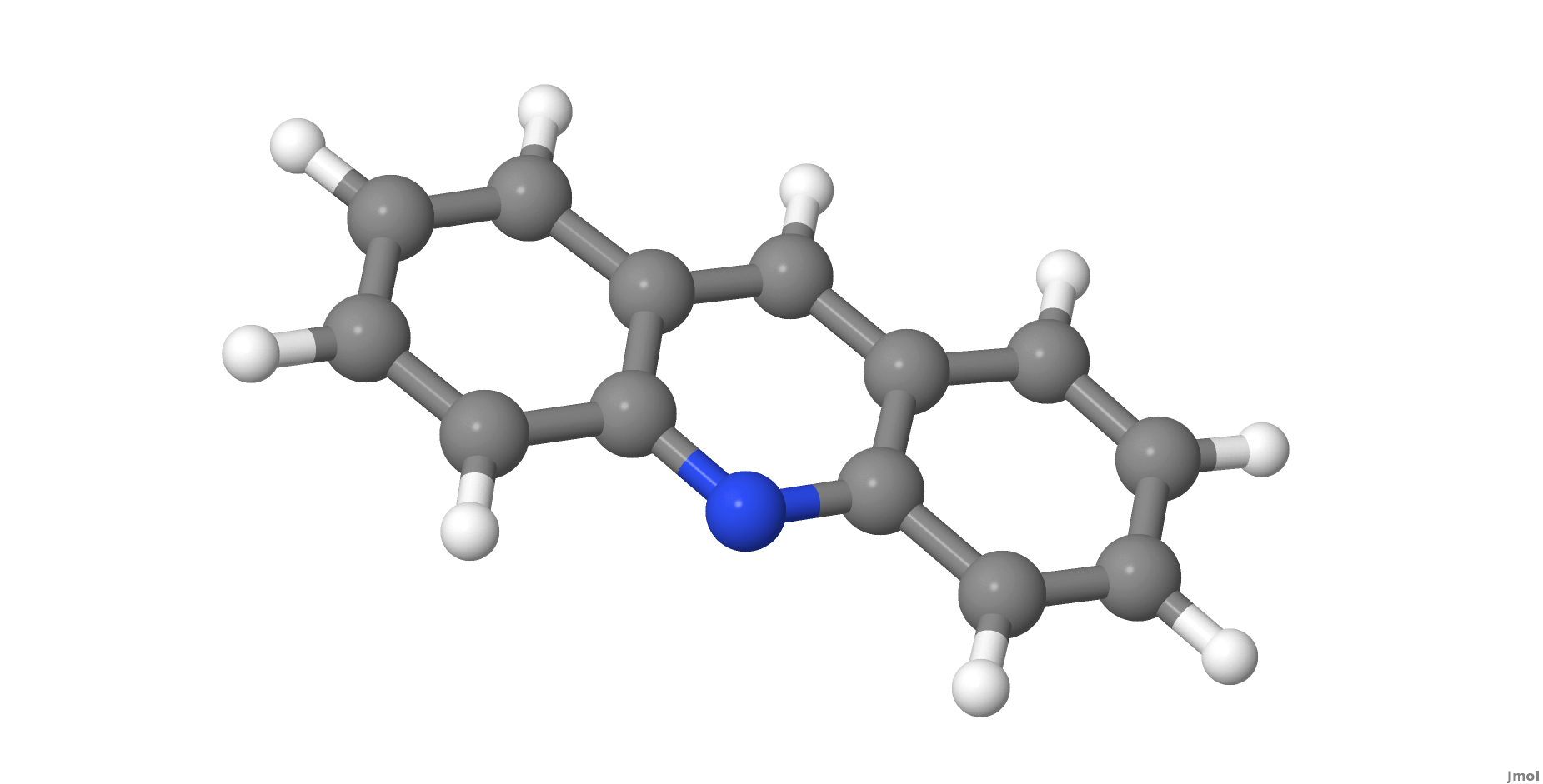}
		\caption{{Acridine}}
	\end{subfigure}
	\begin{subfigure}[b]{0.15\textwidth}
		\includegraphics[scale=0.05]{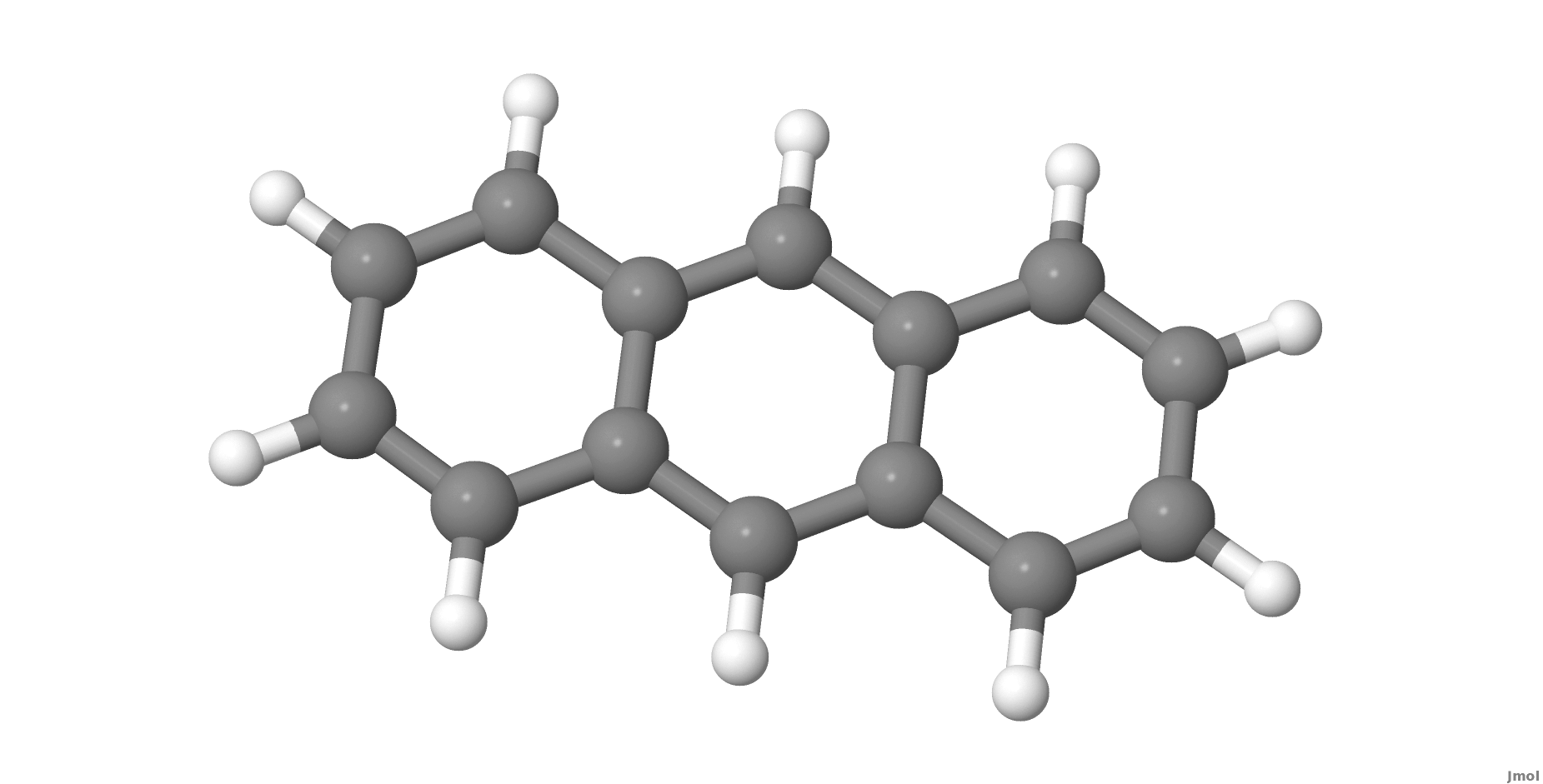} 
		\caption{{Anthracene}}
	\end{subfigure}
	\begin{subfigure}[b]{0.15\textwidth}
		\includegraphics[scale=0.05]{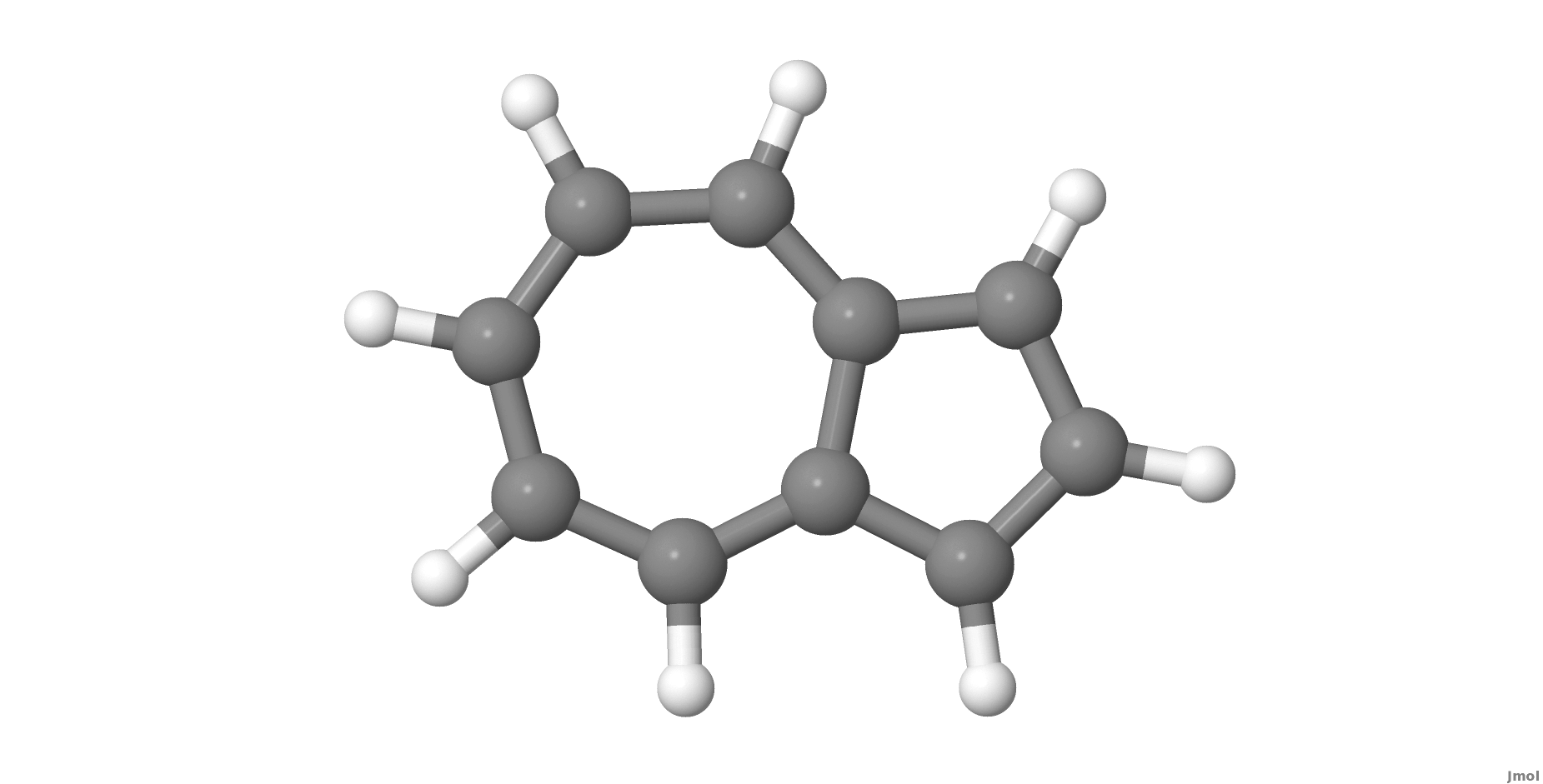} 
		\caption{{Azulene}}
	\end{subfigure}
	\begin{subfigure}[b]{0.15\textwidth}
		\includegraphics[scale=0.05]{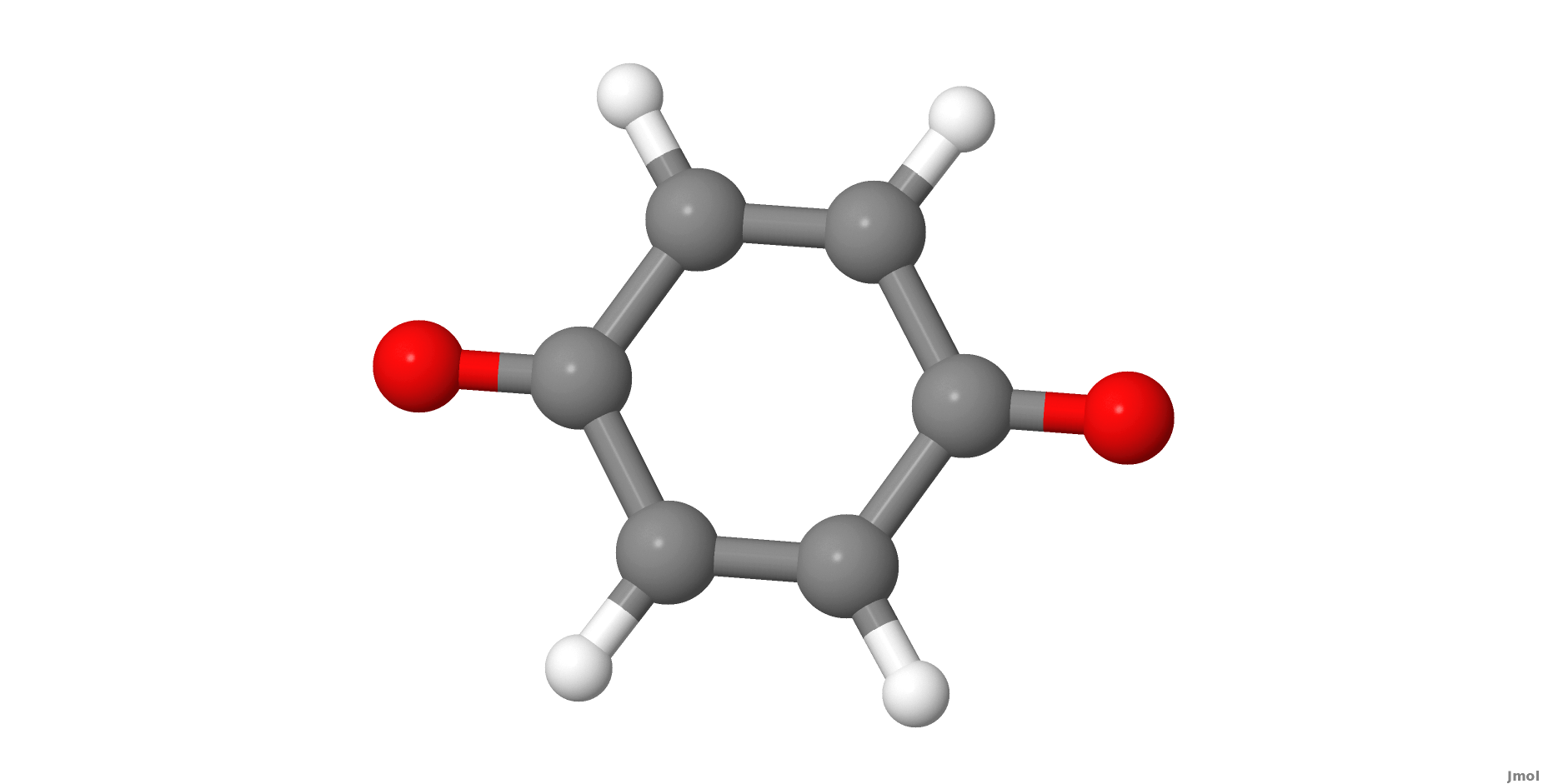}
		\caption{{Benzoquinone(BQ)}}
	\end{subfigure}
	\begin{subfigure}[b]{0.15\textwidth}
		\includegraphics[scale=0.05]{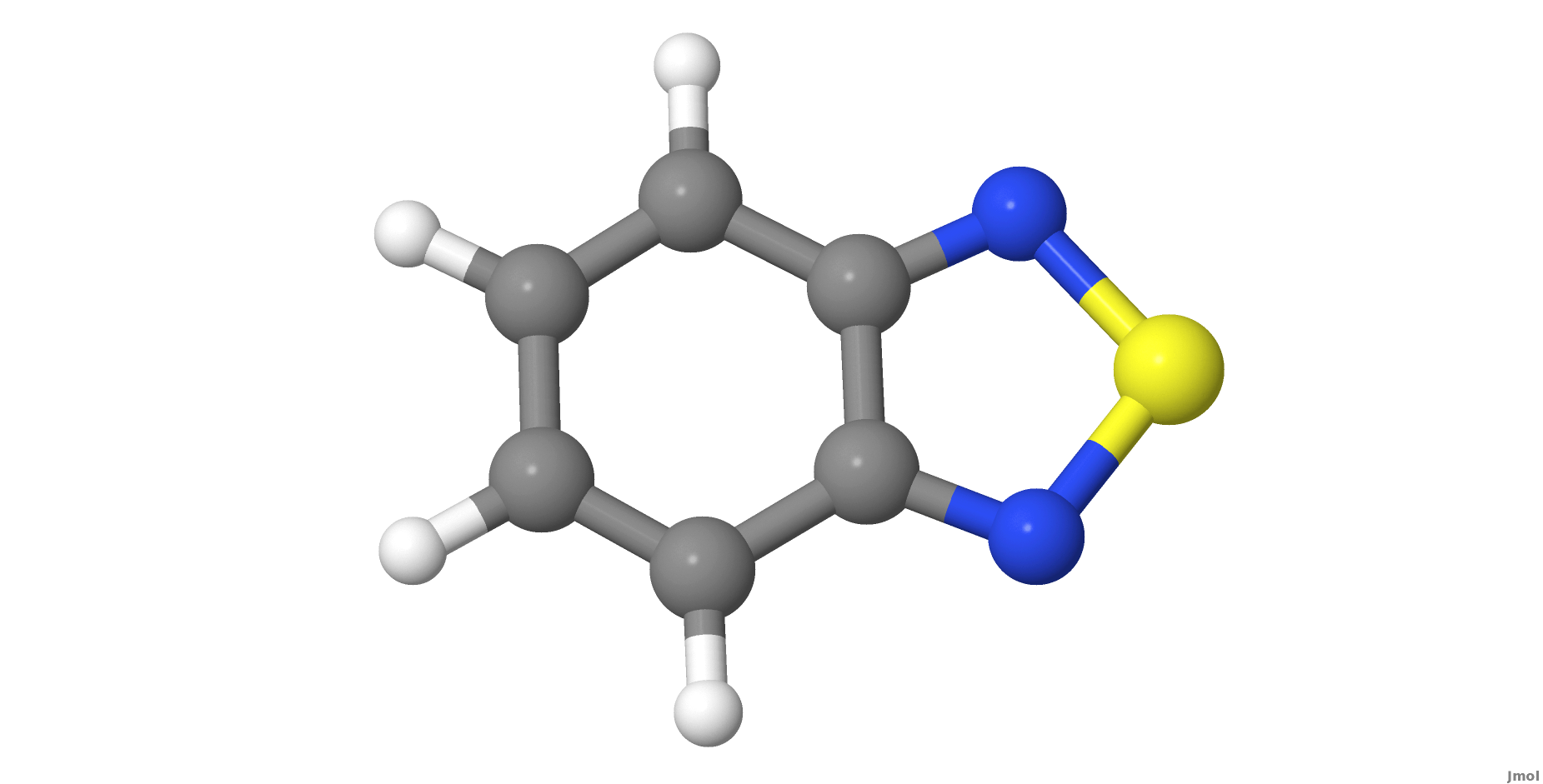}
		\caption{{Benzothiadiazole}}
	\end{subfigure}
	\begin{subfigure}[b]{0.15\textwidth}
		\includegraphics[scale=0.05]{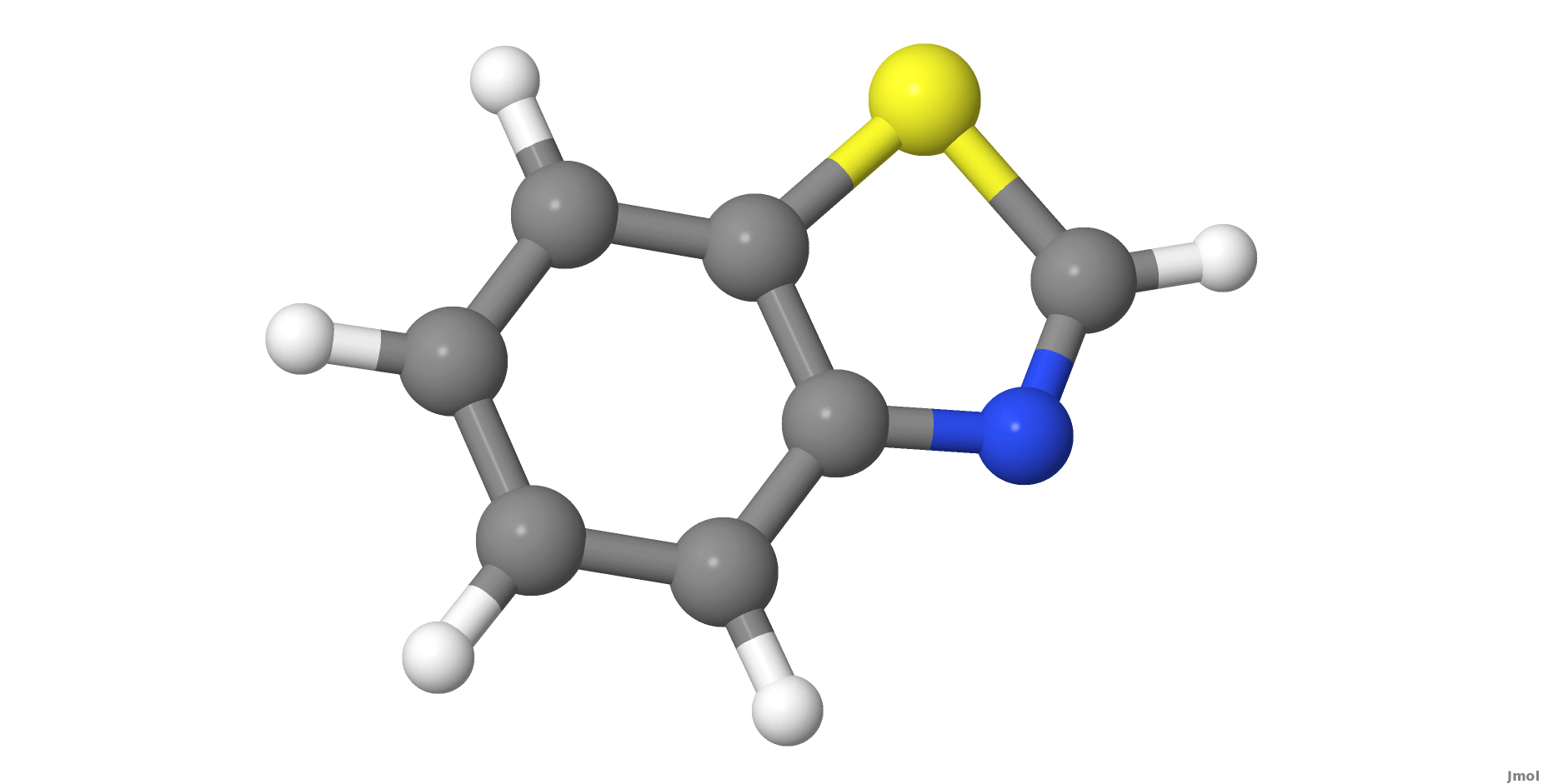} 
		\caption{{Benzothiazole}}
	\end{subfigure}
	\begin{subfigure}[b]{0.15\textwidth}
		\includegraphics[scale=0.05]{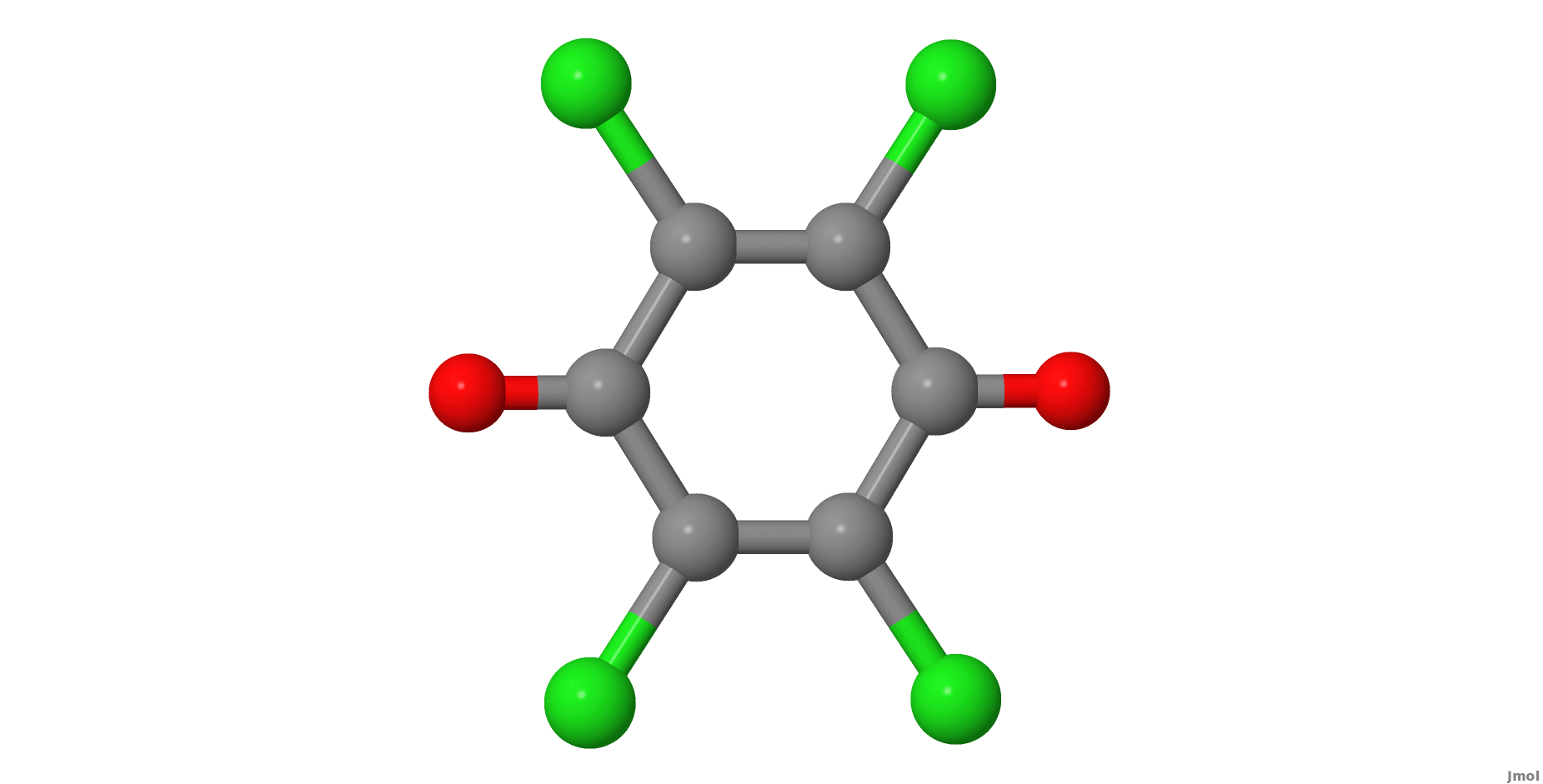} 
		\caption{{Cl4-BQ}}
	\end{subfigure}
	\begin{subfigure}[b]{0.15\textwidth}
		\includegraphics[scale=0.05]{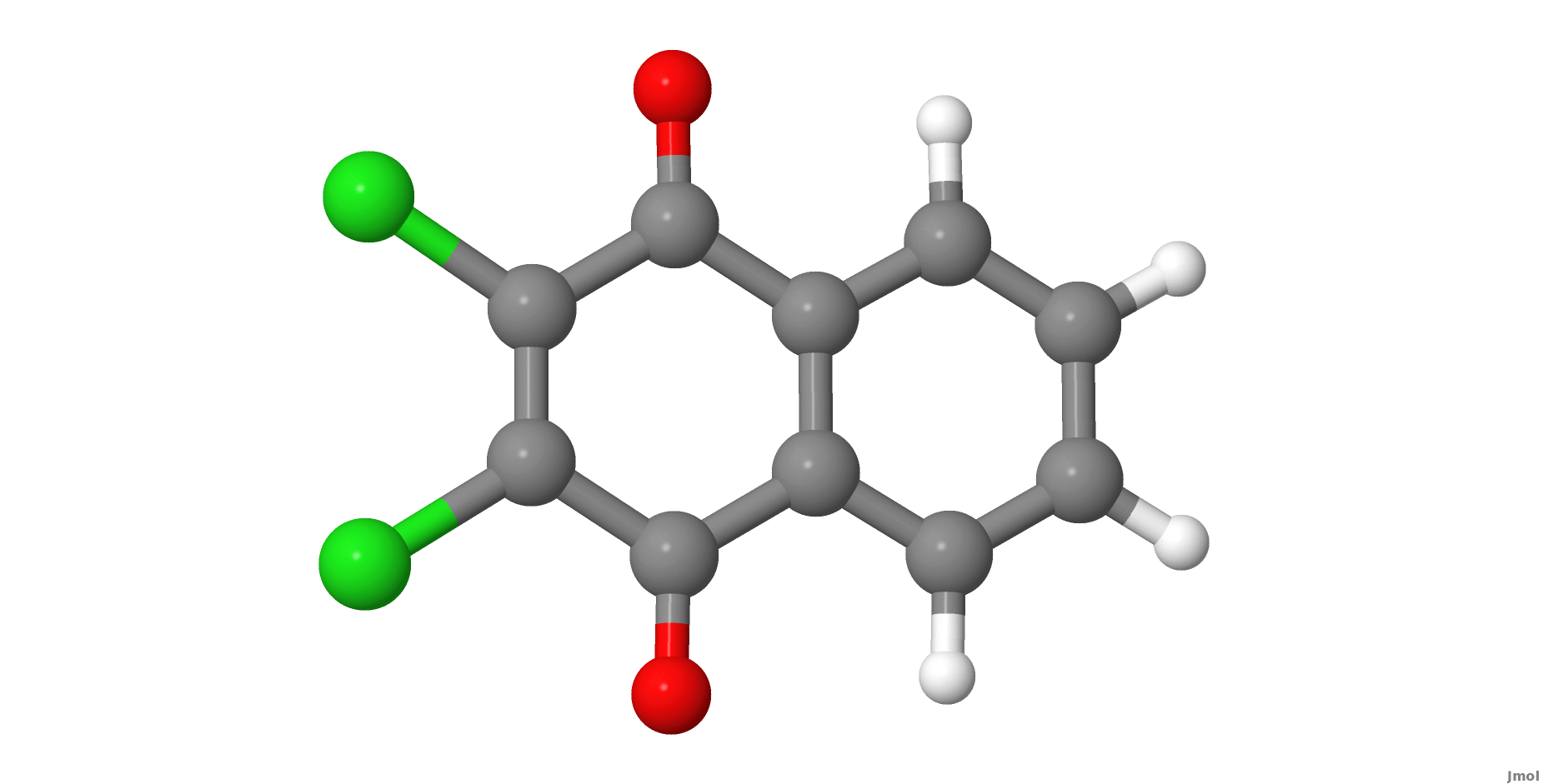} 
		\caption{{Dichlone}}
	\end{subfigure}
	\begin{subfigure}[b]{0.15\textwidth}
		\includegraphics[scale=0.05]{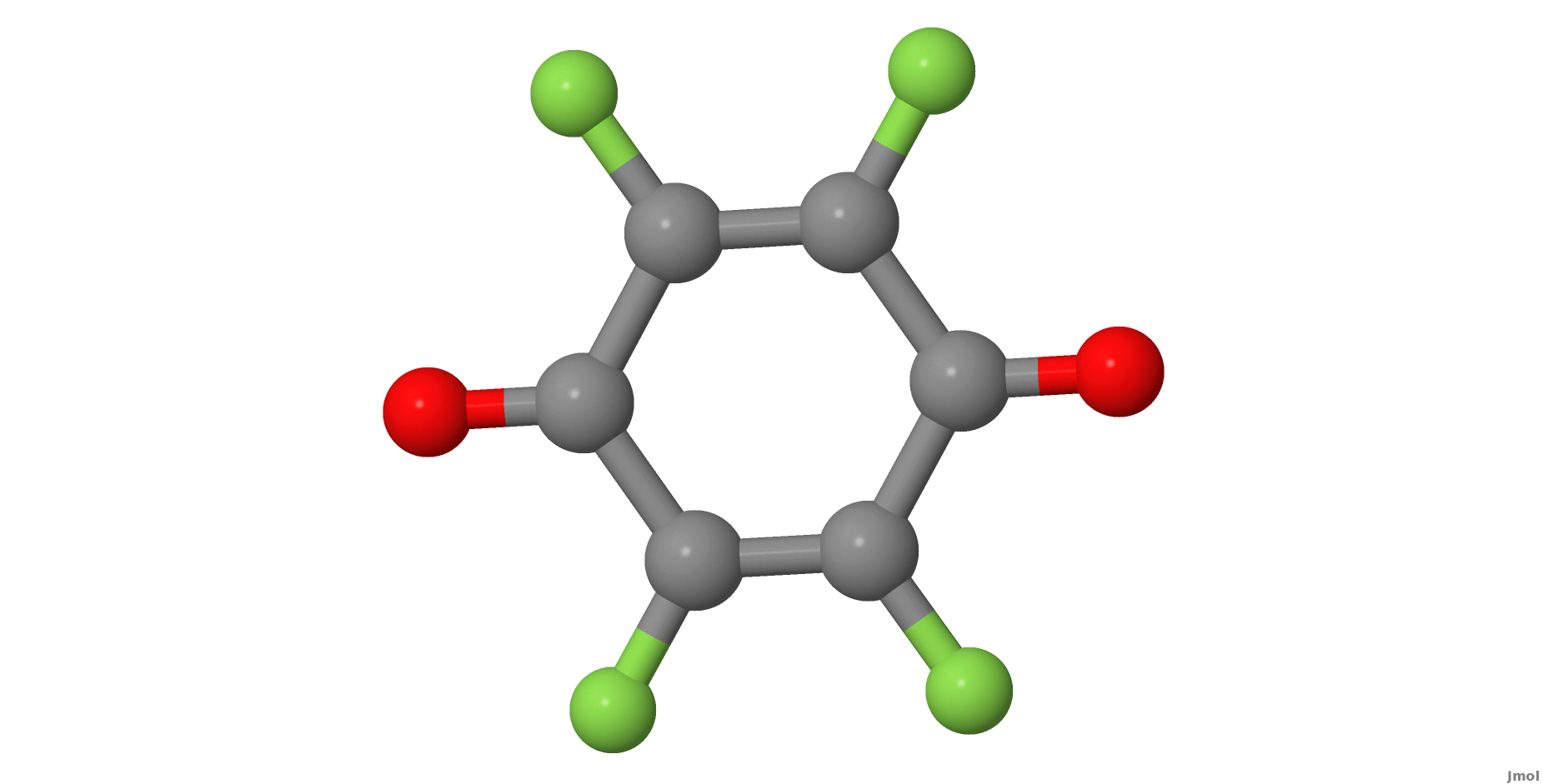} 
		\caption{{F4-BQ}}
	\end{subfigure}
	\begin{subfigure}[b]{0.15\textwidth}
		\includegraphics[scale=0.05]{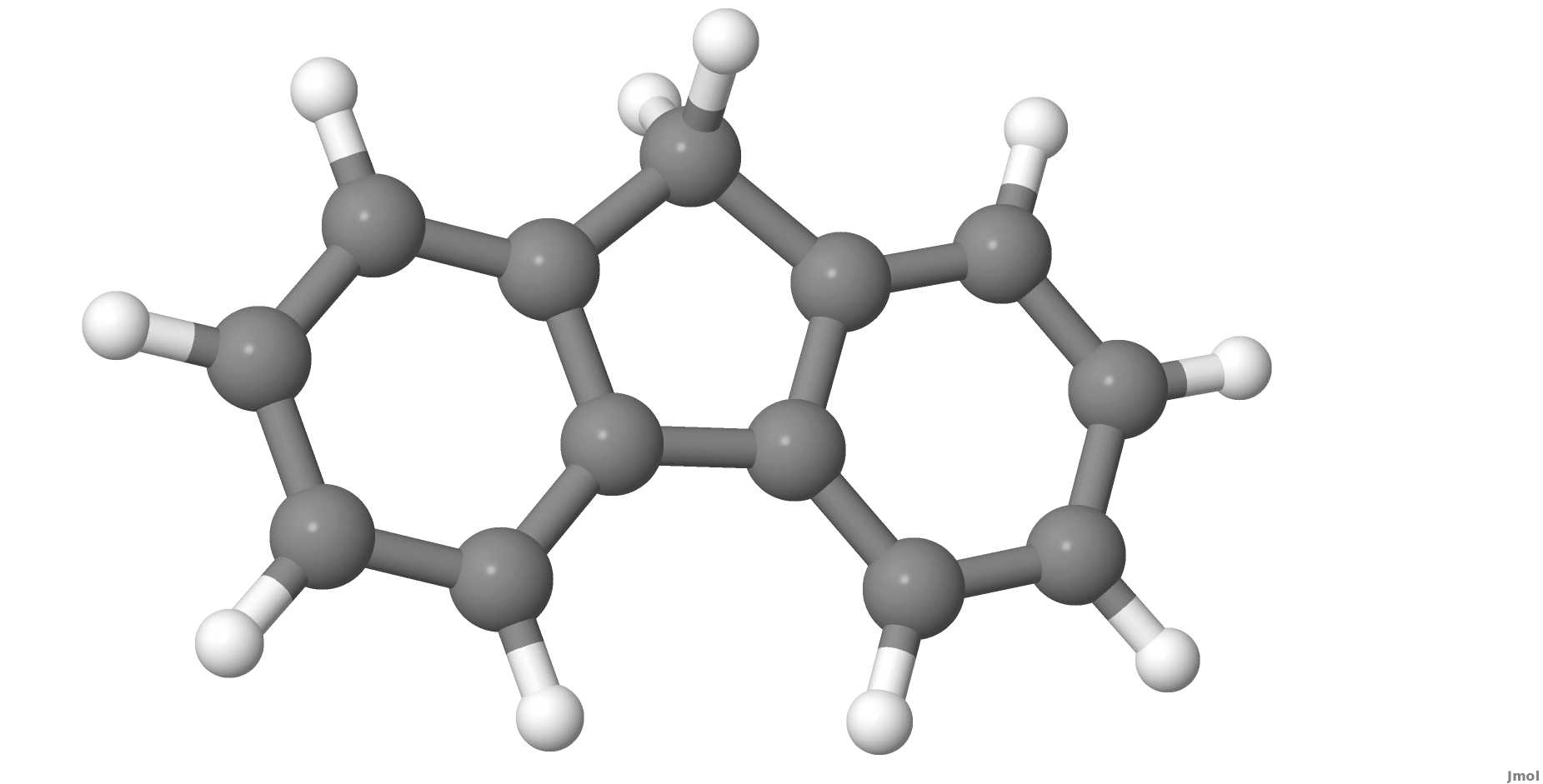}
		\caption{{Fluorene}}
	\end{subfigure}
	\begin{subfigure}[b]{0.15\textwidth}
		\includegraphics[scale=0.05]{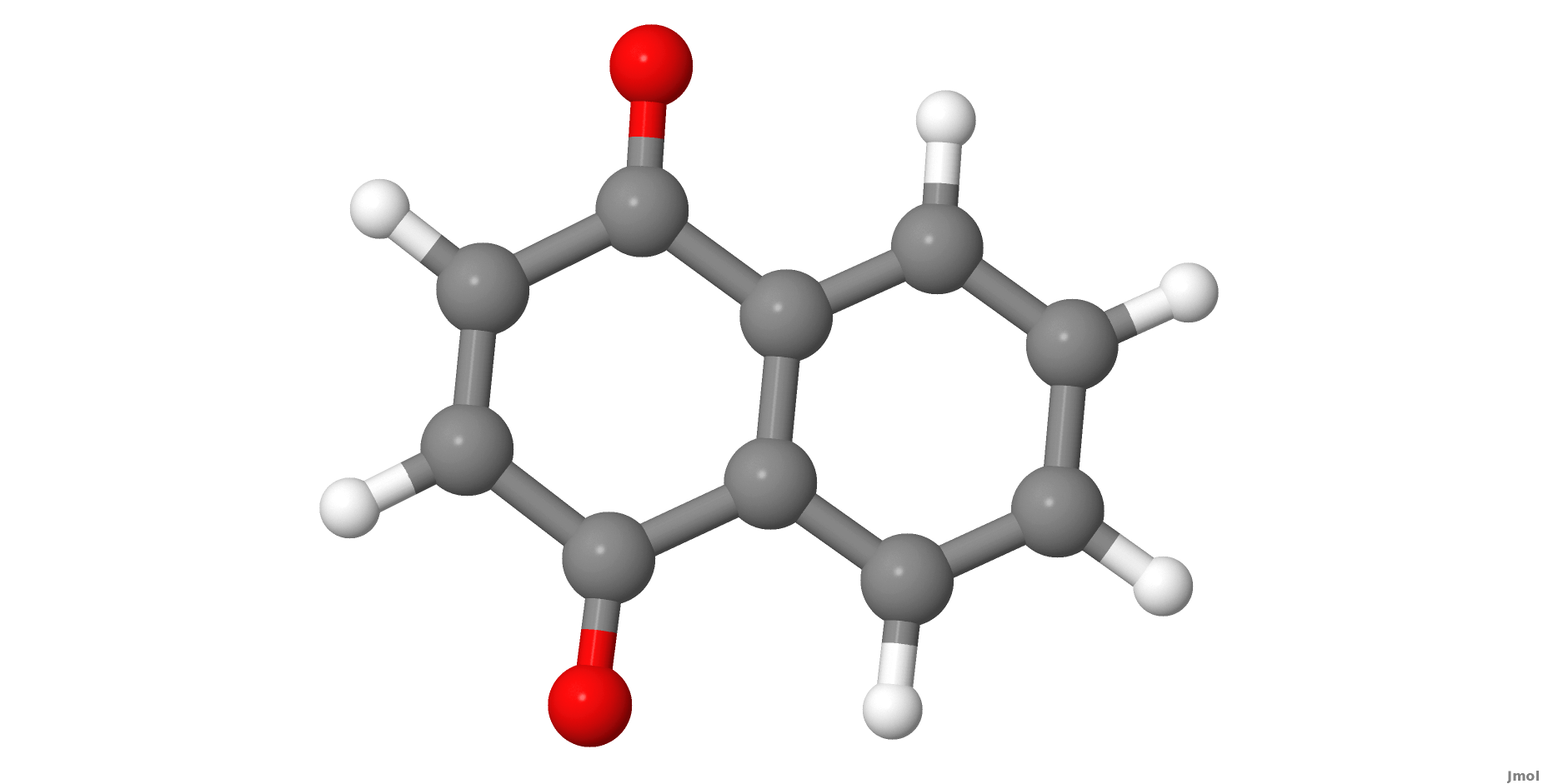} 
		\caption{{Naphthalenedione}}
	\end{subfigure}
	\begin{subfigure}[b]{0.15\textwidth}
		\includegraphics[scale=0.05]{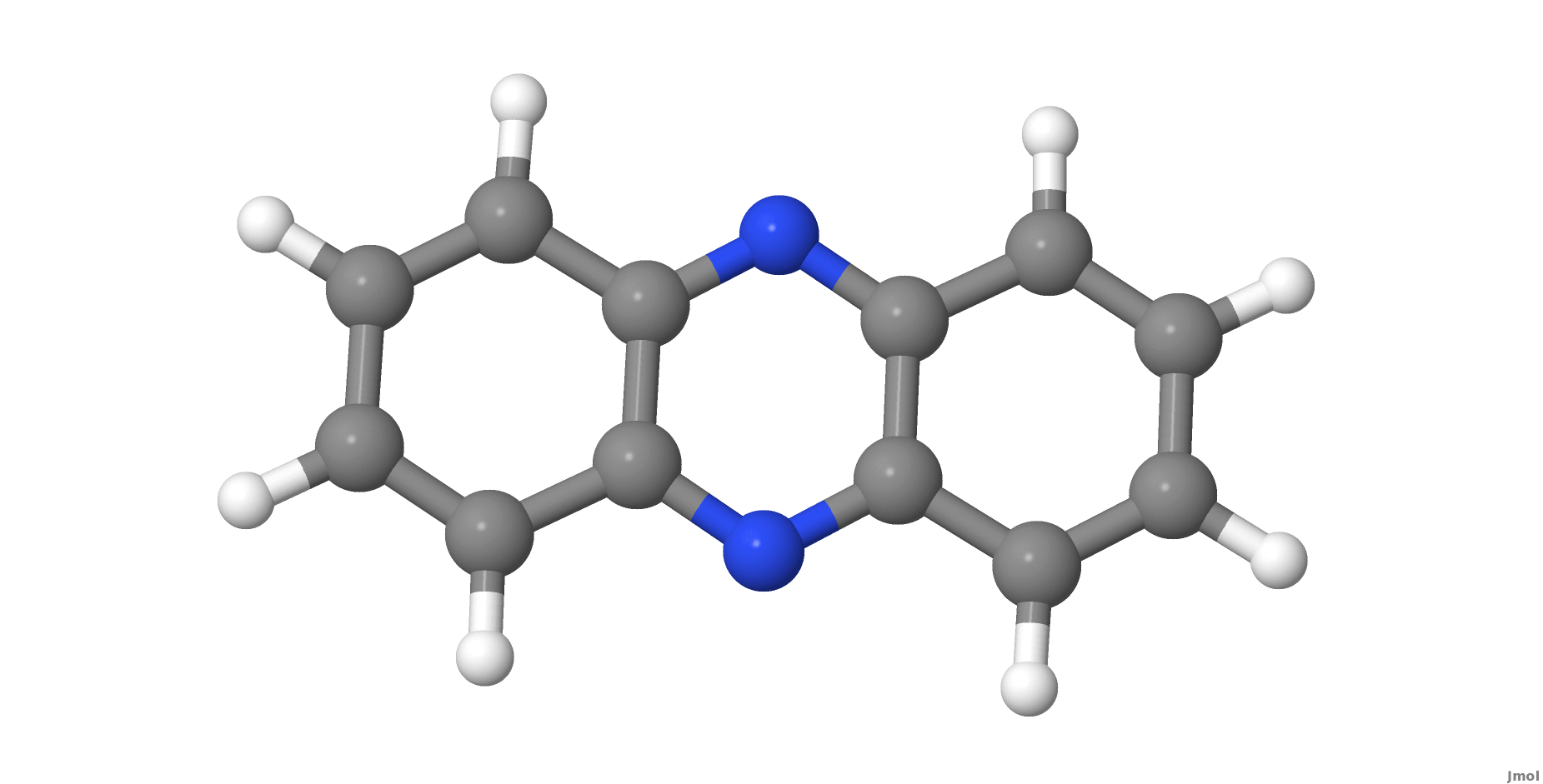} 
		\caption{{Phenazine}}
	\end{subfigure}
	\begin{subfigure}[b]{0.15\textwidth}
		\includegraphics[scale=0.05]{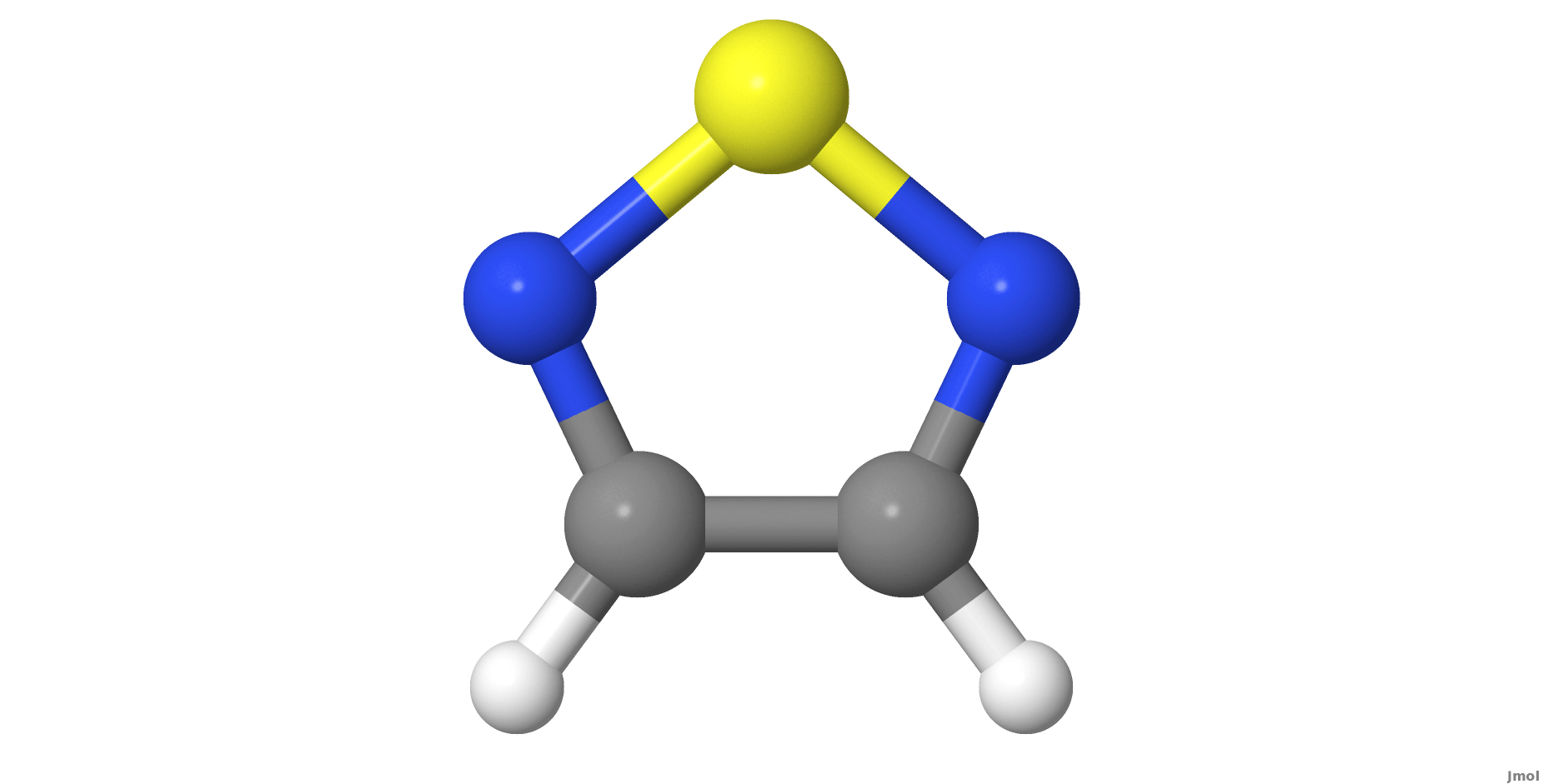} 
		\caption{{Thiadiazole}}
	\end{subfigure}
	\begin{subfigure}[b]{0.15\textwidth}
		\includegraphics[scale=0.05]{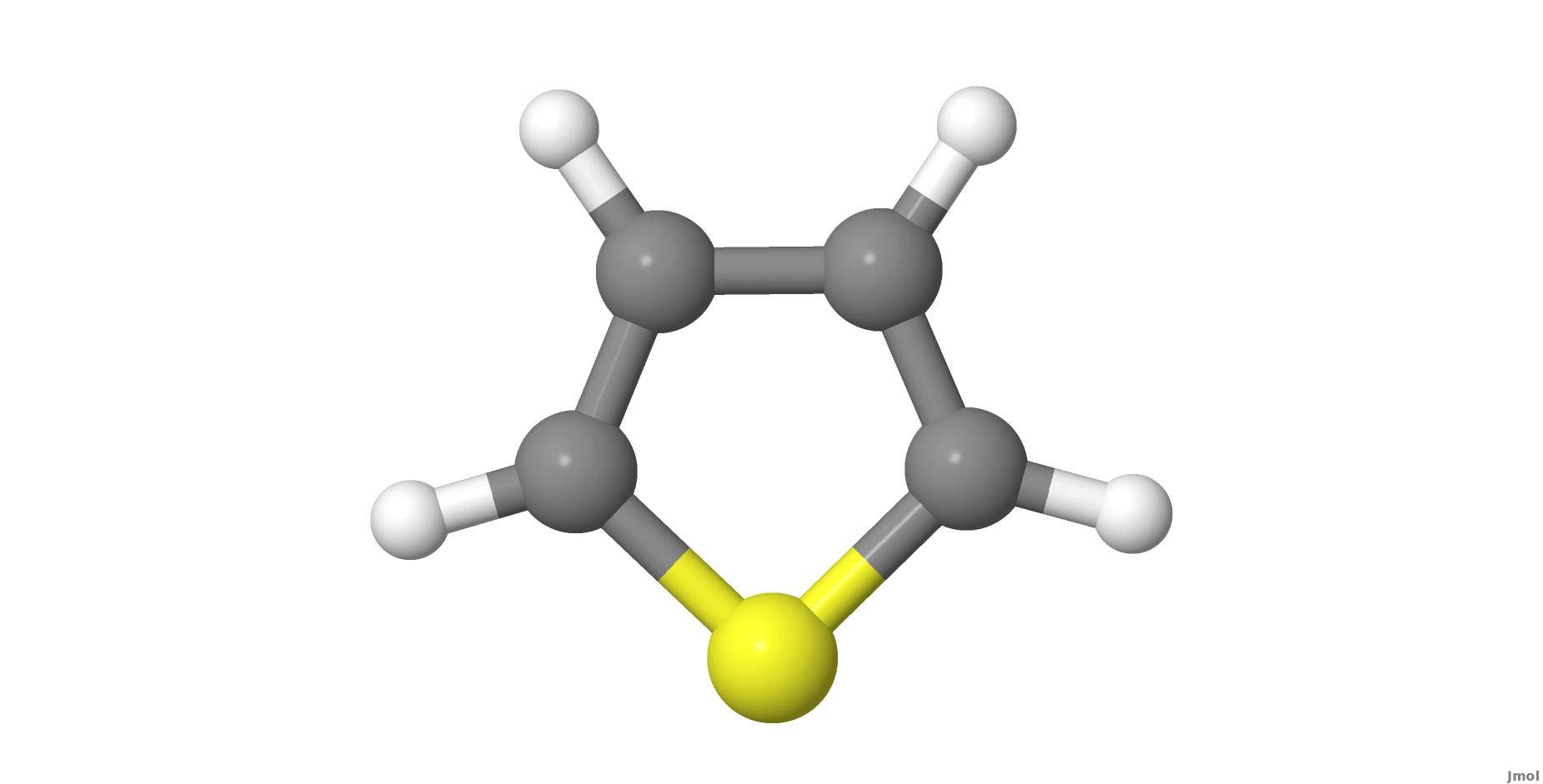} 
		\caption{{Thiophene}}
	\end{subfigure}
	\caption{The geometries of the various organic molecules under study obtained using the software Jmol.\cite{jmol} Small white atoms are hydrogen atoms, gray atoms are carbon atoms, while red, blue, yellow, light green and dark green atoms are oxygen, nitrogen, sulfur, fluorine and chlorine atoms, respectively.}
\end{figure}
$~~~~$Though the exact XC functional would yield the negative eigenvalue of the highest-occupied orbital (HO) as the vertical ionization energy (IE),\cite{perdew1982density,perdew1997comment} it is well known that common DFAs seriously underestimate the magnitude of the HO energies of atoms and molecules. This deviation is attributed to the self-interaction error (SIE) inherent in the approximate functionals and thus one can expect that self-interaction-corrected schemes like FLOSIC can improve the HO energy to near the correct value. There is also an alternative approach to improve the HO eigenvalues obtained from the semilocal approximations through the Koopmans compliant (KC)\cite{dabo2010koopmans,borghi2014koopmans} corrections to the respective semilocal functionals. Such KC functional\cite{dabo2010koopmans} which enforces piecewise linearity (PWL)\cite{perdew1982density} lost in the semilocal density functionals, has been successful for atoms and molecules and even extended systems.\cite{borghi2014koopmans,colonna2019koopmans,nguyen2018koopmans} Since SIE inherent in the semilocal DFAs is responsible for the loss of PWL, self-interaction corrections also help them restore it. So far, the evaluation of IE from FLOSIC has been limited to atoms and small molecules.\cite{schwalbe2018fermi,yamamoto2019fermi} To the best of our knowledge our work is the first to extend self-consistent FLOSIC calculations to the moderate sized organic molecules (shown in Fig. 1). This test set consists of 14 organic molecules which include two important chemical families often used in organic electronics, namely acenes (anthracene, acridine, phenazine and azulene) and quinones (benzoquinone(BQ), F4-BQ, Cl4-BQ, dichlone and naphthalenedione), along with the organic molecules (thiophene, thiadiazole, benzothiazole, benzothiadiazole and fluorene) serving as the building blocks of widely applicable donor polymer materials.\cite{sun2018low} The heterocyclic organic molecules like thiophene and thiadiazole have possible applications in catalysis\cite{wang2010thiophene,adhikari2020} and as corrosion inhibitors.\cite{loto2012corrosion,guo2018anticorrosive} In the first part of our work we evaluate the IEs of our test set using regular FLOSIC methods and compare them with experimental IEs available in the NIST database.\cite{nistwebbook} For completeness, we also perform non-self-consistent GW calculations as a theoretical reference. In the second part of our work we discuss the performance of a recently developed local-scaling SIC scheme\cite{zope2019step} on our test set, along with the performance of the newly introduced scaling schemes discussed in the Methodology section. Throughout this work we will be evaluating IE as the negative of the HO energy. 
\section{Methodology}
Despite the exactness of PZ-SIC for one electron systems, it has a tendency to overcorrect many-electron systems.\cite{vydrov2004effect,vydrov2005ionization} As a remedy, Vydrov \textit{et al.} made an effort to scale down PZ-SIC in the many-electron regime using an exterior scaling scheme involving a single adjustable parameter.\cite{vydrov2006scaling,vydrov2006simple} The scaled-down version proved to be superior over PZ-SIC for properties like atomization energies, barrier heights, bond lengths etc. but, unlike PZ-SIC, it failed to predict the correct dissociation of the neutral molecules such as NaCl.\cite{ruzsinszky2006spurious} Furthermore the scaled-down version was shown to deviate more than PZ-SIC does from the agreement of vertical IE with the negative of the highest occupied orbital eigenvalue, a condition that an exact functional should satisfy.\cite{perdew1982density,perdew1997comment} Kl\"upfel \textit{et al.}\cite{klupfel2012effect} proposed a global scaling down of the SIC terms in Eq. (3) by a factor of one half. Despite not working well with LDA, such scaling when applied with PBE slightly improved the atomization energies of small molecules compared to full SIC, However, barrier heights were not improved by this scaling. Recently, Zope \textit{et al.}\cite{zope2019step} introduced a scheme to scale the SIC energy density at each point in space using an iso-orbital indicator ($z_{\sigma}$(\textbf{r})). This quantity, whose value lies between 0 and 1, is the ratio of the von Weizs\"acker kinetic energy density (${\tau}_{\sigma}^{W}(\textbf{r})$) to the Kohn-Sham kinetic energy density (${\tau}_{\sigma}(\textbf{r})$) (as shown in Eq. (5)).
\begin{equation} 
z_{\sigma}(\textbf{r})=\frac{{\tau}_{\sigma}^{W}(\textbf{r})}{{\tau}_{\sigma}(\textbf{r})}
\end{equation}
where,
\begin{equation*}
	{\tau}_{\sigma}^{W}(\textbf{r})=\frac{{|{\nabla}n_{\sigma}(\textbf{r})|}^2}{8n_{\sigma}(\textbf{r})}
\end{equation*}
and
\begin{equation*}
	{\tau}_{\sigma}(\textbf{r})=\frac{1}{2}\sum_{i=1}{|{\nabla}{\psi}_{i\sigma}(\textbf{r})|}^{2}.
\end{equation*}
This scheme, known as local-scaling SIC (LSIC), distinguishes the one-electron density where a full SIC is applied from the uniform density where no correction is applied. The latter is also the limit where PZ-SIC is known to deviate from the behavior of the exact functional.\cite{santra2019perdew}
This approach re-evaluates the correction terms in the RHS of Eq. (3) using Eqs. (6) and (7) utilizing the scaling function $f(z_{\sigma})$ of Eq. (8).
\begin{equation}
U[n_{i\sigma}] = \frac{1}{2}\int d\textbf{r}~f(z_{\sigma})n_{i\sigma}(\textbf{r})\int \frac{{n_{i\sigma}(\textbf{r$'$})}}{|\textbf{r}-\textbf{r$'$}|}~d\textbf{r$'$}
\end{equation}

\begin{equation}
E^{app}_{xc}[n_{i\sigma},0]=\int d\textbf{r}~f(z{_\sigma}) n_{i{\sigma}}(\textbf{r}){\epsilon}_{xc}^{app}([n_{i{\sigma}}(\textbf{r}),0],\textbf{r})
\end{equation}
\begin{equation}
f^{LSIC}(z_{\sigma}) = {z_{\sigma}}.
\end{equation}
This scaling\cite{zope2019step} applied to LDA (LDA-LSIC), significantly improves the description of both equilibrium and non-equilibrium properties. 

\begin{figure}[h!]
	
	\includegraphics[scale=0.5]{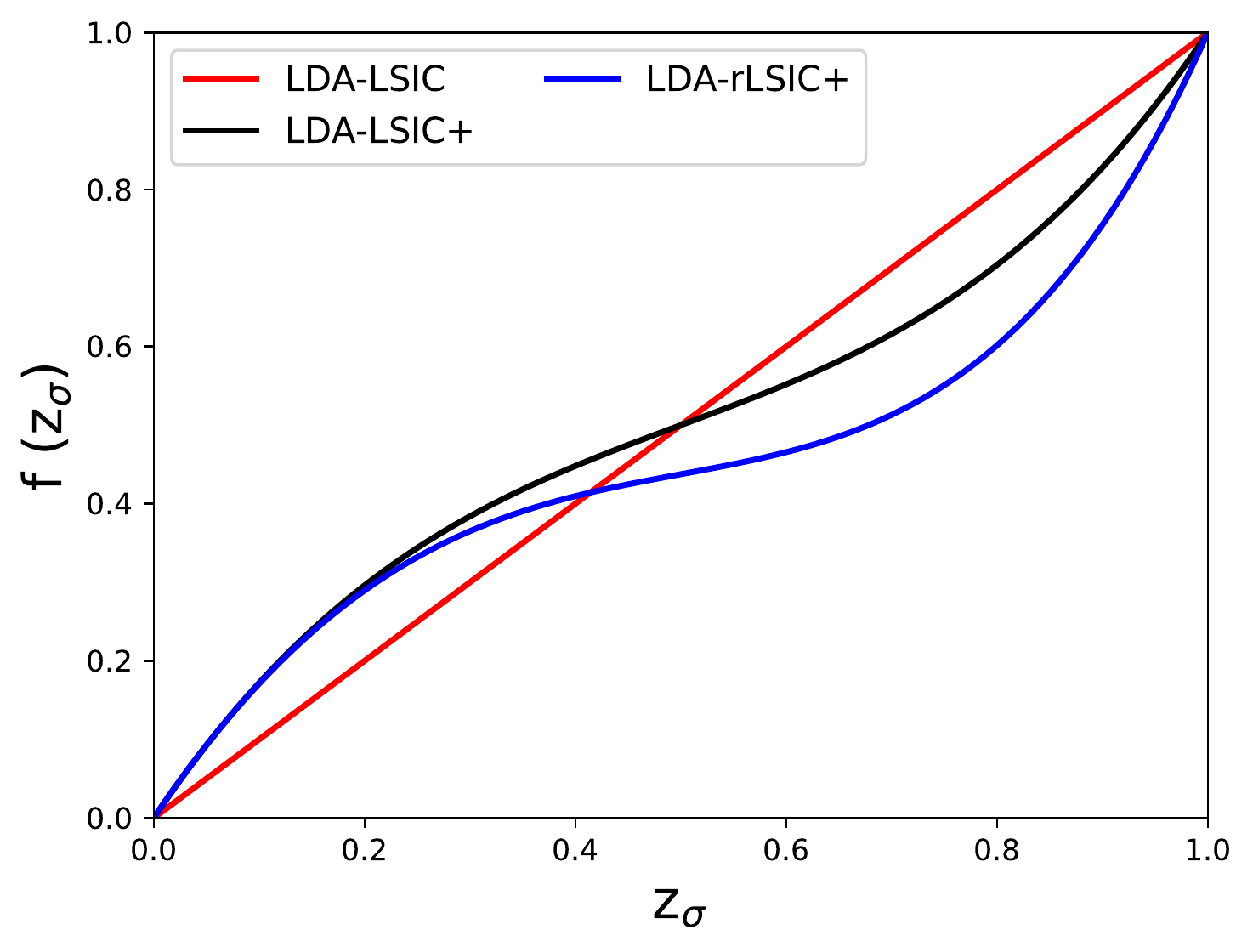} 
	
	\caption{ The plot of f($z_{\sigma}$) vs $z_{\sigma}$ for different scaling schemes. {While f($z_{\sigma}$) = 0 and f($z_{\sigma}$) = 1 reduce to LDA and LDA-SIC respectively}, all these approaches guarantee no correction for the uniform density limit and full correction at single density limit. }
\end{figure} 
The LDA-LSIC scaling function employs a linear interpolation between the one-electron density limit ($z\to1$) and the uniform electron density limit ($z\to0$). Linear interpolation is a straightforward choice which leads to a significant improvement over PZ SIC.\cite{zope2019step} However, LDA-LSIC is not necessarily optimial for all many-electron density regions.\cite{bhattarai2020step} To go beyond LSIC, here we use two new scaling schemes namely LDA-LSIC+\cite{bhattarailsicplus} and its revised version, LDA-rLSIC+, using the scaling functions as shown in Eqs. (9) and (10) respectively.
\begin{equation}
f^{LSIC+}(z_{\sigma}) = a + a(z_{\sigma} - a) + 4(1 - a){(z_{\sigma} - a)}^{3},~~a = 0.5,
\end{equation}

\begin{multline}
	f^{rLSIC+}(z_{\sigma}) = a + (8a - 3){(z_{\sigma} - a)} -3(3a - 1){(z_{\sigma} - a)^2} \\ + (4a + 1){(z_{\sigma} - a)^3} + {(z_{\sigma} - a)^4},~~a=\sqrt{2} - 1.
\end{multline}	
Both these methods assign ${z_{\sigma}}$-dependent corrections to the corresponding energy densities and are devised only for LDA to avoid gauge inconsistency.\cite{bhattarai2020step} LDA-LSIC+\cite{bhattarailsicplus} is specifically designed to reduce the significant error of LDA-SIC for neutral atoms in the limit of large atomic numbers.\cite{santra2019perdew} Compared to LDA-LSIC, LDA-LSIC+ is more correcting towards the  ${z_{\sigma} \to 0}$ region, while less towards the ${z_{\sigma} \to 1}$. We found the IEs of the molecules in our test set improved by LDA-LSIC+ compared to LDA-LSIC but to achieve significant improvement the scaling function of LDA-LSIC+ needed to be revised. The revised version of LDA-LSIC+ (LDA-rLSIC+) is even less correcting than LDA-LSIC+ towards the ${z_{\sigma} \to 1}$ region. The constant "a" in Eqs. (9) and (10) is introduced to distinguish such regions (see appendix A). The value of a = $0.5$ in the scaling function of LDA-LSIC+ indicates that the scaling function is more correcting than LDA-LSIC in the region 0 < ${z_{\sigma}}$ < 0.5, less correcting in the region 0.5 < ${z_{\sigma}}$ < 1.0, and reduces to LDA-LSIC at ${z_{\sigma}}$ = 0.5. Similar conclusion follows for $a = \sqrt(2) - 1$ for LDA-rLSIC+. As a consequence of the modification, LDA-rLSIC+ is not as accurate as LDA-LSIC+ for neutral atoms in the limit of large atomic numbers.  Specifically, the error in the leading coefficient of the large-Z asymptotic expansion of the exchange-correlation energy obtained with LDA-SIC (5.4$\%$) is largely reduced when computed with LDA-LSIC (-0.6$\%$), LDA-LSIC+ (0.6$\%$), and LDA-rLSIC+ (1.0$\%$). These large-Z limit errors are obtained by extrapolating errors from Ne, Ar, Xe, Kr (as shown in Fig. 3 of Santra \textit{et al.}\cite{santra2019perdew}). The root-mean-squared-percentage-error for the four rare-gas atoms is largely reduced from LDA-SIC (3.7$\%$) to LDA-LSIC (0.6$\%$), LDA-LSIC+ (0.4$\%$), and LDA-rLSIC+ (0.9$\%$). Importantly, the second coefficient in the large-Z asymptotic expansion is improved towards the exact value of -0.187\cite{burke2016locality,cancio2018fitting} in both LDA-LSIC+ (-0.183) and LDA-rLSIC+ (-0.196), as opposed to LDA-LSIC (-0.085) and LDA-SIC (-0.724).\cite{bhattarailsicplus} Nevertheless, all these scaled-down methods guarantee no correction for a uniform electron density (${z_{\sigma}}$=0) and full corrections for a single electron density (${z_{\sigma}}$=1). Figure 2 displays a plot of the scaling function $f(z_{\sigma})$ as a function of $z_{\sigma}$.
We note in passing that we do not attempt to apply local-scaling in the case of PBE. A primary reason is that the PBE energy densities are not in the same gauge of the Hartree energy and stratightforward local-scaling destroys the balance between these two energy densities.\cite{bhattarai2020step} LDA and Hartree energy densities are in the same gauge allowing for different variants of local-scaling of the PZ SIC.
\section{Computational details}
The initial geometries of all the molecules were obtained from the NIST database\cite{nistwebbook} and were not further optimized in any of the computations. Non-self-consistent G$_0$W$_0$ calculations based on PBE and Hartree-Fock (HF) references were performed using the all-electron numerical atom-centered orbital (NAO) code, FHI-aims.\cite{blum2009ab,ren2012resolution} The default \textit{tier 4} basis set proven to predict quasi-particle energies converged to within 0.1 eV\cite{knight2016accurate} was used for these calculations. The FLOSIC code based on the UTEP version of the NRLMOL code\cite{flosic} was used for all other calculations. 
Spin unpolarized calculations using the FLOSIC code were performed on these closed shell molecules using the Gaussian type NRLMOL default basis set.\cite{porezag1999optimization} The self-consistent convergence criteria on total energy was set to 10$^{-7}$ Ha and while the geometry was kept fixed, the Fermi orbital descriptors (FODs) were optimized. The trial FODs required to initiate the calculations were generated using the Monte-Carlo method\cite{21} based on minimizing the Coulomb repulsion (fodMC). The FOD optimization was performed until the largest component of an FOD force was less than 5 X 10$^{-4}$ Ha/Bohr, and the energy difference between the two successive steps was less than 10$^{-6}$ Ha. We have performed fully-self-consistent calculations  with LDA-SIC and PBE-SIC.\cite{yang2017full} 
\section{Results and discussion}
\subsection{Ionization Energy (IE)}
In the first part of the work we have assessed the vertical IEs of our test set comprising 14 small to medium sized organic molecules using LDA and PBE along with their self-interaction corrected counterparts (denoted by LDA-SIC and PBE-SIC, respectively). Figure 3 displays the overall performance of the various methods. Our calculations show LDA seriously underestimating the IEs and this underestimation becomes slightly worse with PBE. Although self-interaction corrected LDA and PBE improve the IEs, they largely overcorrect them. The mean error of LDA, PBE and their SIC counterparts computed with respect to the experimental values obtained from the NIST database,\cite{nistwebbook} as shown in Fig. 4, are large compared to our non-self-consistent GW calculations. We have performed non-self-consistent GW calculations using the ground-state eigenstates and eigenvalues from both PBE (G$_0$W$_0$@PBE) and HF (G$_0$W$_0$@HF), and found the former method to be slightly superior over the latter. The details of these calculations can be found in the Supplementary Tables S1 and S2. Despite the fact that self-interaction corrections systematically improve the uncorrected functionals, the overall performance is still far from satisfactory.
\begin{figure}[h!]	
	\includegraphics[scale=0.6]{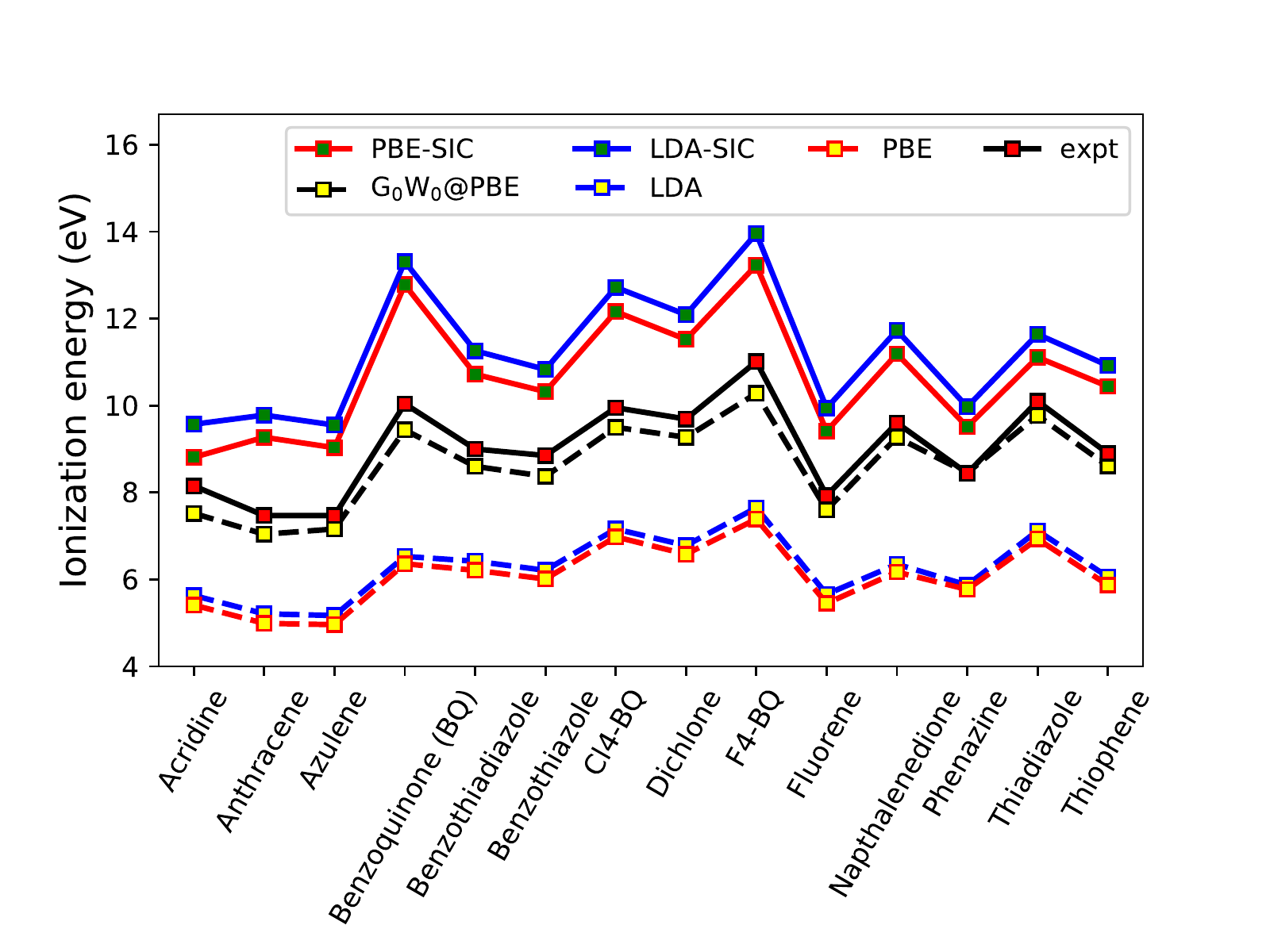} 
	\caption{Ionization energy evaluated as the negative of HO eigenvalue using different methods. We can clearly see LDA and PBE underestimating the IE while their SIC counterparts are systematically overestimating.}
\end{figure} 
\begin{figure}[h!]
	\includegraphics[scale=0.5]{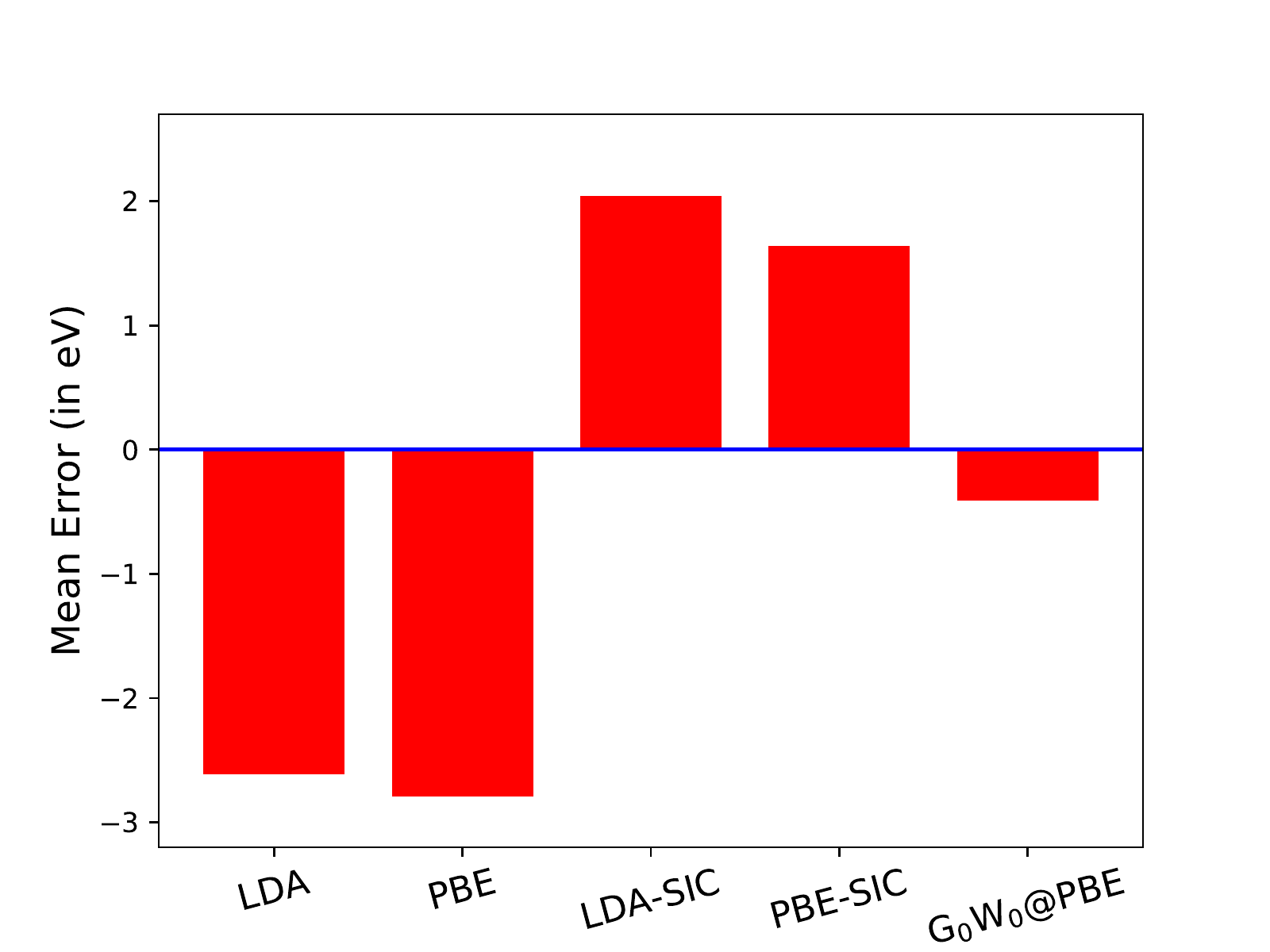} 
	\caption{Mean error (in eV) when ionization energy of the molecules in our test set is evaluated as the negative of the HO eigenvalue using LDA, PBE and their SIC counterparts.}
\end{figure}

Our test set consisting of organic molecules is a paradigm of delocalized many-electron systems, but unfortunately, past studies\cite{vydrov2004effect,vydrov2005ionization} have shown PZ-SIC to be overcorrecting IEs in many-electron systems. Recent work on the IE of atoms and small molecules\cite{schwalbe2018fermi,yamamoto2019fermi} using the FLOSIC method has displayed such overestimation, too. Thus overcorrection in many-electron systems is the consequence of PZ-SIC being designed to be exact only for all one-electron systems. The other serious problem of PZ-SIC is that it loses\cite{santra2019perdew} exactness in the uniform electron density limit, preserved by all the non-empirical functionals.\cite{perdew2005prescription} PZ-SIC thus overcorrects the region of organic molecules where the electron density approximates the uniform density limit. Thus the need to develop a method which makes correction only in the region where SIE is important is inevitable.   \\
$~~~~$In the second part of the work we have assessed the IEs of our test set using various scaled-down schemes designed to simultaneously fix the overcorrection in many-electron regions and preserve exactness in the one-electron and uniform density limits. Figures 5 and 6 display the overall performance of these methods. Earlier we showed LDA-SIC being superior to LDA for evaluating IE, but still the predicted values were far from the experimental ones. However, with the scaling schemes, there is a remarkable improvement over the performance of LDA-SIC. The LDA-LSIC scheme of Zope \textit{et al.} \cite{zope2019step} reduces the ME to 1.13 eV which is significantly better than the 2.04 eV of LDA-SIC. The other scaling schemes, namely LDA-LSIC+ and LDA-rLSIC+, designed to weight the corrections more or less heavily in different regions of $z_{\sigma}$, perform even better. With a ME of just 0.40 eV, the LDA-rLSIC+ method performed on par with the theoretical reference method, G$_{0}$W$_{0}@$PBE. The details of the calculations can be found in the Supplementary Table S3.\\
\begin{figure}[h!]
	
	\includegraphics[scale=0.6]{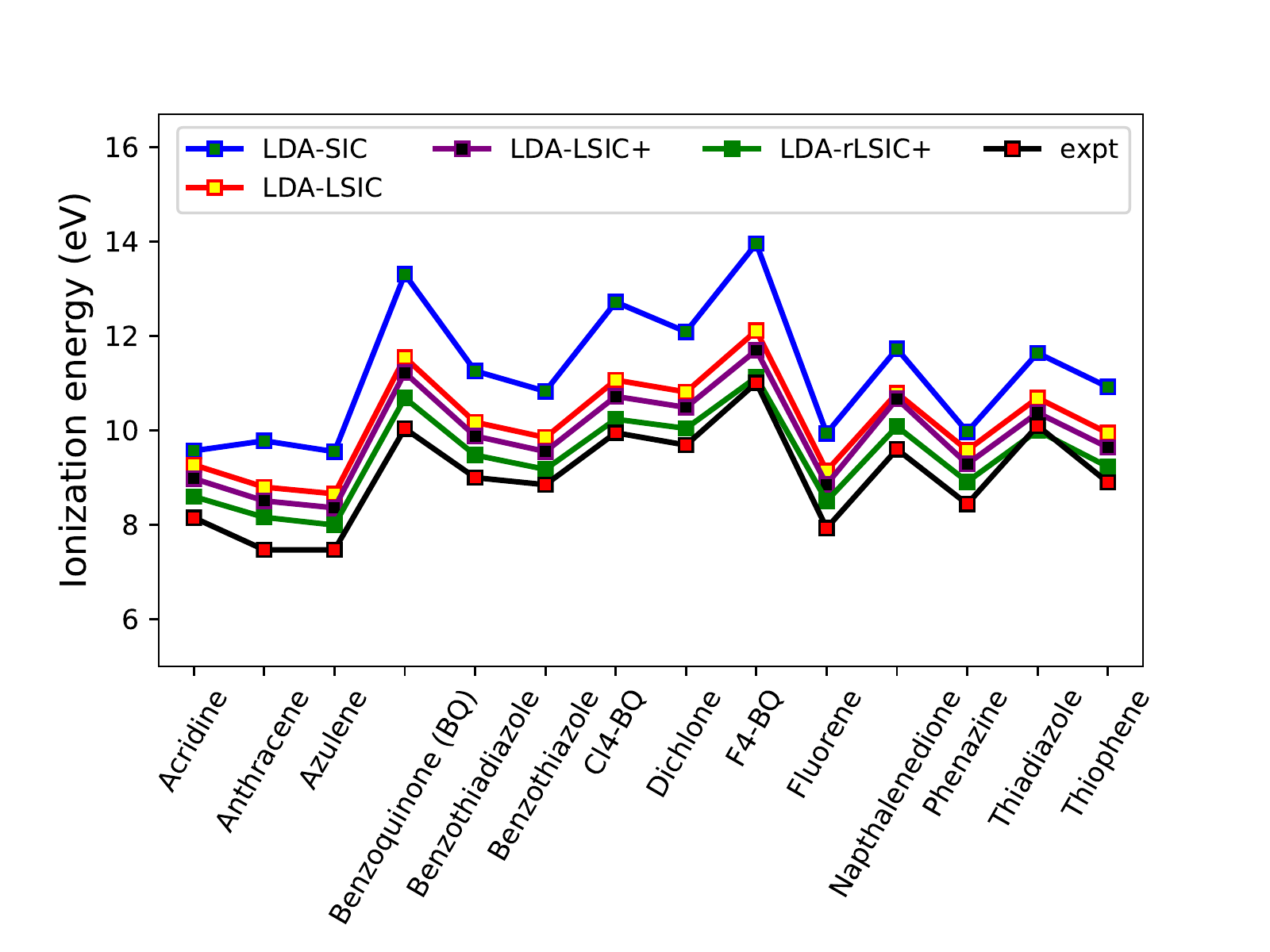} 
	
	\caption{Ionization energy as the negative of HO eigenvalue using different scaling schemes. All the scaled-down approximations perform significantly better than LDA-SIC. The LDA-rLSIC+ approach in particular stands out. }
\end{figure}  
\begin{figure}[h!]
	
	\includegraphics[scale=0.5]{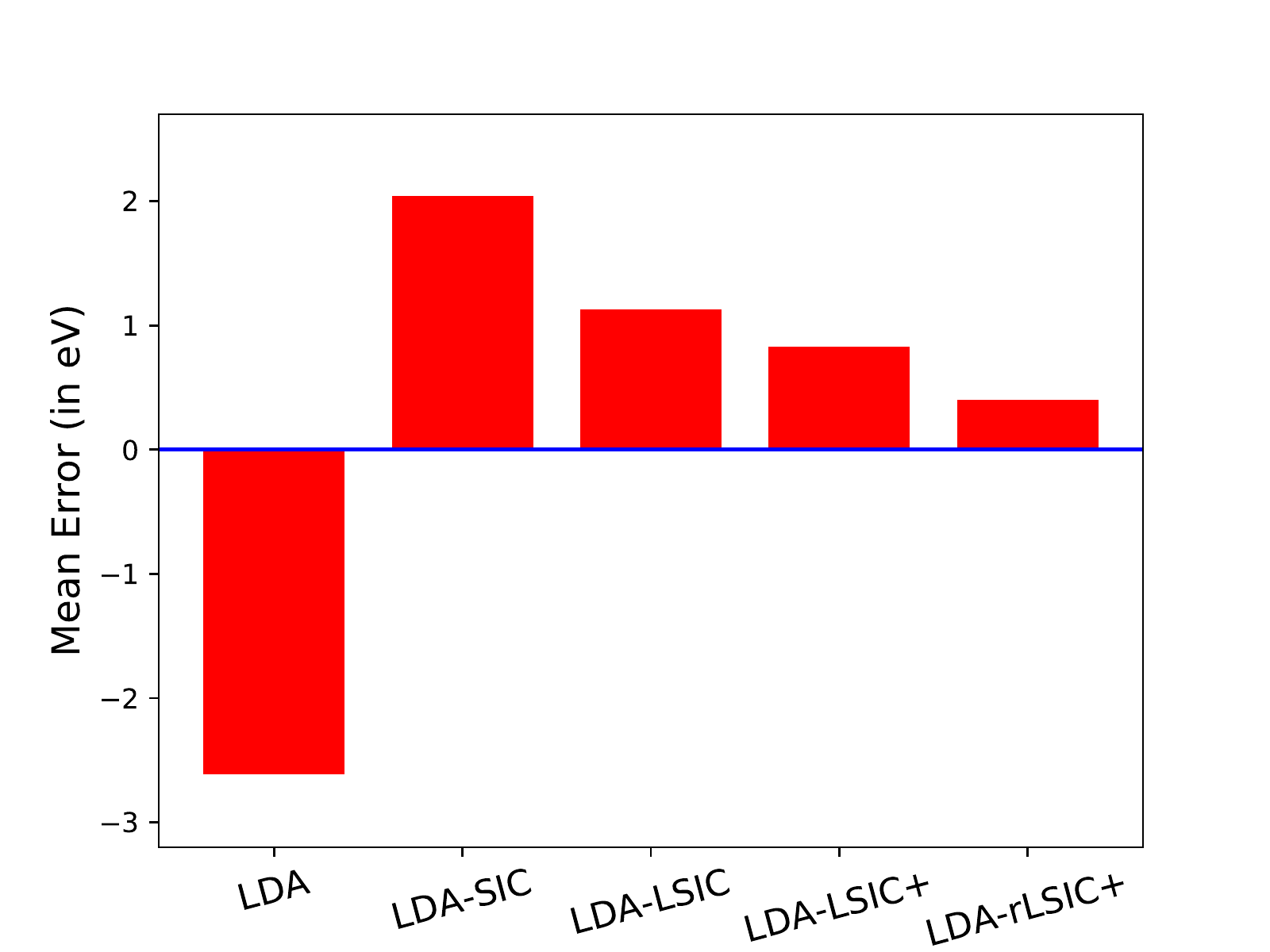} 
	
	\caption{Mean error (in eV) of the ionization energy of the molecules in our test set evaluated as the negative of the HO using using different scaling schemes. The LDA-rLSIC+ scaling approach performs on par with the theoretical reference method. }
\end{figure}  
We also evaluated the IEs of the G2-1 test set\cite{curtiss1991gaussian} using these various scaled-down approaches. The details of these calculations can also be found in the Supplementary Table S4. Figure 7 displays the ME and MAE of the various scaling methods along with LDA and LDA-SIC. Here again the scaling approaches stand out, performing significantly better than LDA-SIC. The performance of LDA-rLSIC+ for the G2-1 test set is not as impressive as it was for the organic molecules, but, while not better, it almost performs at the level of the other two scaling methods. LDA-LSIC+ method performs slightly better than the other two scaling approaches for the G2-1 test set. \\
\begin{figure}[h!]
	
	\includegraphics[scale=0.5]{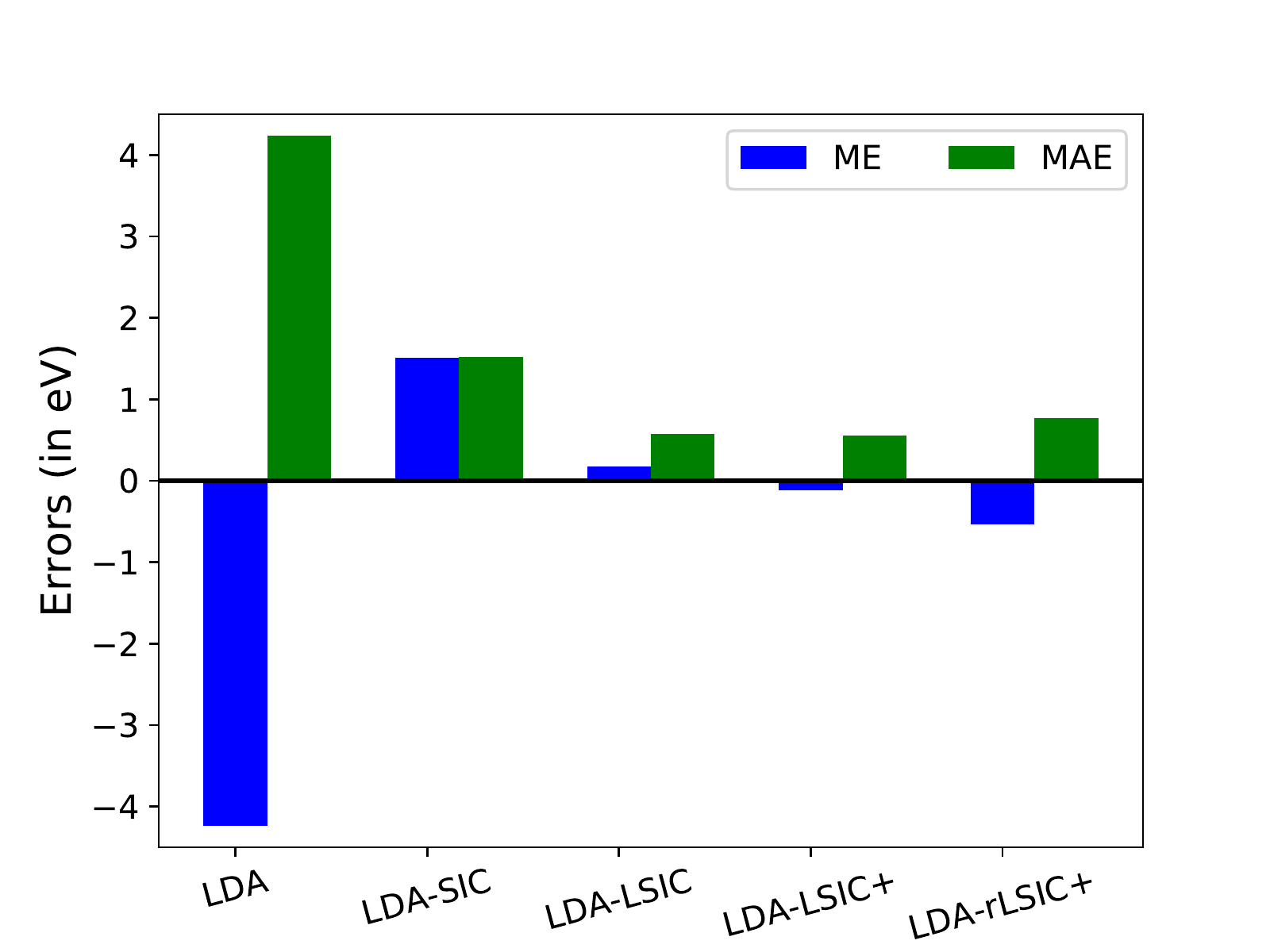} 
	
	\caption{ME and MAE (in eV) when IE of G2-1 test set is evaluated as the negative of the HO using using different scaling schemes.  }
\end{figure}  

\subsection{Dipole Moments}
The dipole moment measures the charge distribution within a system and distinguishes whether a molecule is polar or non-polar. Experimental dipole moments\cite{livermore1964tables} of organic molecules are usually measured using a solvent (mostly benzene as it is non-polar) at a finite temperature. We have assessed and compared the dipole moments obtained from DFT calculations that correspond to values in the gas phase and at absolute zero temperature with the experimental dipole moments whenever available.  Based on our results shown in Supplementary Table S5, the (semi)-local functionals LDA, PBE and SCAN without SIC give a better accuracy for the dipole moments. Both LDA-SIC and PBE-SIC worsen the calculated dipole moments. The scaling schemes introduced here significantly improve the dipole moments that are made worse by regular SIC and are almost as accurate as the uncorrected semi-local methods. An assessment of the dipole moments of a molecular test set by Withanage \textit{et al.}\cite{dipolemomentKushantha} exhibits significant improvement of the scaled-down methods over the regular FLOSIC methods too. The scaled-down methods even outperform LDA and PBE for assessment of dipole moments of that molecular test set but are not as accurate as SCAN. The ME and MAE analysis of dipole moments of our test set shows that the uncorrected semi-local functionals perform very close to each other as LDA, PBE and SCAN have almost the same MAE. The same trend follows for fully corrected and scaled-down functionals. Due to this reason we just show the error of LDA, LDA-SIC and LDA-rLSIC+ in Fig.8 as the representatives of the three different classes. The details of the calculations for all the molecules using different methods can be found in the Supplementary Table S5. \\
\begin{figure}[h!]
	
	\includegraphics[scale=0.5]{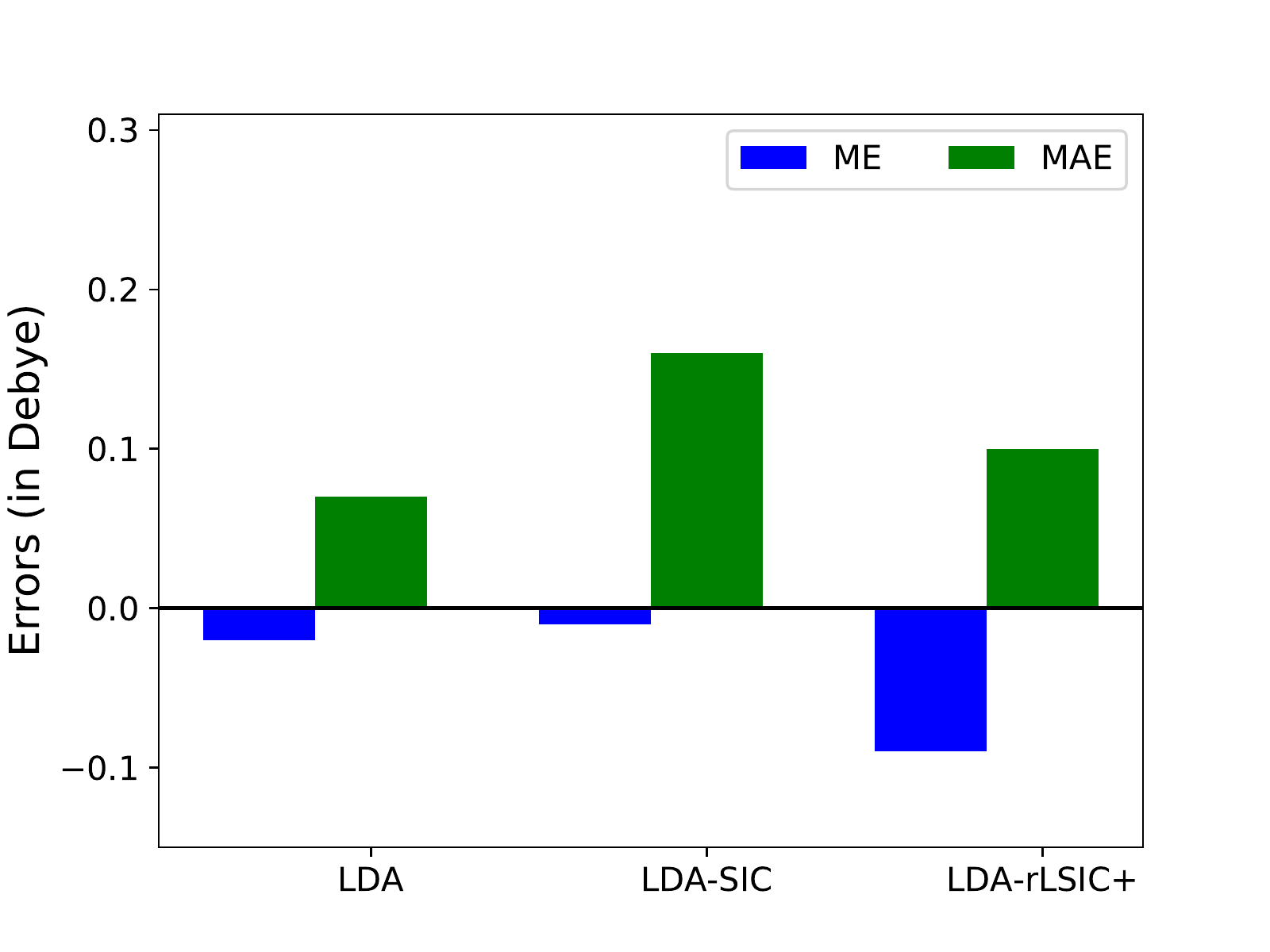} 
	
	\caption{ME and MAE of the dipole moment (in Debye) of the organic molecules in our data set obtained using LDA, LDA-SIC, and LDA-rLSIC+.}
\end{figure}  
\section{Conclusion}
We have evaluated the ionization energy (IE) of 14 small to moderate-sized organic molecules, as the negative of the highest-occupied orbital (HO) eigenvalue, using various self-interaction correction (SIC)-based approximations to density functionals. We used the local-scaling SIC (LDA-LSIC) scheme of Zope \textit{et al.}\cite{zope2019step} and introduced different scaling schemes, namely LDA-LSIC+\cite{bhattarailsicplus} and LDA-rLSIC+, based on different scaling functions of the iso-orbital indicator ($z_{\sigma}$). We have shown that the latter schemes which weight the corrections more or less heavily than the original LDA-LSIC in different regions of $z_\sigma$ perform better for the IEs of our test set. \\
We also evaluated the IEs of the G2-1 test set\cite{curtiss1991gaussian} using these scaling schemes and confirmed the importance of the scaled-down SIC on these systems, too. For completeness, we also extended this assessment to the dipole moments of our test set and found the scaled-down methods to significantly improve over the unscaled methods, performing almost at the level of the uncorrected functionals.\\
There is a possibility of further improvement by imposing a full self-consistency in the LSIC schemes including the functional derivatives of the $z_{\sigma}$ iso-orbital indicator. The impact of full self-consistency will be investigated in a future work. However, we expect this to have a minor effect and would not diminish the performance of the scaling methods.

\begin{acknowledgements}
	The work of S.A., B.S., A.R., and K.A.J. was supported by the U.S Department of Energy, Office of Science, Office of Basic Energy Sciences, as part of the Computational Chemical Sciences Program under Award No. DE-SC0018331. The work of N.K.N. and S.R. was supported by the National Science Foundation under the grant number DMR-1553022. The work of P.B. was supported by the National Science Foundation under the grant number DMR-1939528. This research includes calculations carried out on HPC resources supported in part by the National Science Foundation through major research instrumentation grant number 1625061 and by the US Army Research Laboratory under contract number W911NF-16-2-0189. We thank Prof. John P. Perdew , Biru KC, Kamal Wagle, Dr. Kai Trepte , Kushantha P.K. Withanage and Dr. Rajendra Joshi for their valuable help during the calculations.	\\~\\
	
\end{acknowledgements}
\begin{appendix}
	\noindent
	\textbf{APPENDIX A: ABOUT EQS. (9) AND (10)\\}
	$~~$The simplified form of the scaling function of LDA-LSIC+ (Eq. (9)) is
	\renewcommand{\theequation}{A1}
	\begin{equation}
	f^{LSIC+}(z_{\sigma}) = {2z_{\sigma}}-3{z_{\sigma}^2}+{2z_{\sigma}^3}.
	\end{equation}
	At the value of $z_{\sigma} ~=~ 0.5$, the plot of LDA-LSIC+ and LDA-LSIC intersect separating the regions $z_{\sigma} ~<~ 0.5$ where LDA-LSIC+ is more correcting than LDA-SIC from the region $z_{\sigma} ~>~ 0.5$ where LDA-LSIC+ is less correcting than LDA-LSIC. Expressing Eq. (A1) about a point a = 0.5, we obtain 
	\renewcommand{\theequation}{A2}
	\begin{equation}
	f^{LSIC+}(z_{\sigma}) = a + a(z_{\sigma} - a) + 4(1 - a){(z_{\sigma} - a)}^{3},~~a = 0.5.
	\end{equation}
	Eq. (A2) is same as Eq. (9).\\
	LDA-rLSIC+ was designed to be less correcting than LDA-LSIC+ in the region of $z_{\sigma} \to $ 1 by introducing fourth order term of $z_\sigma$ in the scaling function as
	\renewcommand{\theequation}{A3}
	\begin{equation}
	f^{rLSIC+}(z_{\sigma}) = {2z_{\sigma}}-3{z_{\sigma}^2}+{z_{\sigma}^3}+{z_{\sigma}^4}.
	\end{equation}
	The plot of LDA-rLSIC+ scaling function intersects that of the LDA-LSIC in the region of $z{_\sigma}$ between 0 and 1 at $\sqrt{2} - 1$.
	Setting a = $\sqrt{2}-1$ and expanding the scaling function of LDA-rLSIC+ about "a", we obtain 
	\renewcommand{\theequation}{A4}
	\begin{multline}
		f^{rLSIC+}(z_{\sigma}) = a + (8a - 3){(z_{\sigma} - a)} -3(3a - 1){(z_{\sigma} - a)^2} \\ + (4a + 1){(z_{\sigma} - a)^3} + {(z_{\sigma} - a)^4}.
	\end{multline}	
	Eq. (A4) is same as Eq. (10).\\~\\
	
\end{appendix}
%

\end{document}